\documentclass[11pt,lot, lof]{puthesis}

\usepackage{graphicx,psfrag}
\usepackage{amsfonts}
\usepackage{amssymb}
\usepackage{amsmath}
\usepackage{epsfig, color}
\usepackage{setspace}
\usepackage{epstopdf}
\usepackage{bm}

\newcommand{\proquestmode}{}

\title{Various probes of Dirac matter: from graphene to topological insulators \\(Part I)}

\submitted{29 November 2012}  
\copyrightyear{2012}  
\author{J\'{e}r\^{o}me Cayssol}
\adviser{Des Rapporteurs:\\ \medskip Karyn Le Hur, Laurent Levy, Pascal Simon. \medskip\\Et des Examinateurs:\\  \medskip Alexandre Buzdin, Benoit Dou\c{c}ot, \\Daniel Est\`{e}ve, Gilles Montambaux et \\Bernard Pla\c{c}ais}  
\department{compos\'{e}}

\usepackage{graphicx}

\usepackage{verbatim}

\usepackage{multirow}
\usepackage{longtable}

\usepackage{booktabs}

\ifdefined\printmode

\usepackage{pdfpages}
\usepackage{url}

\else

\ifdefined\proquestmode

\usepackage{hyperref}

\hypersetup{bookmarksnumbered}

\makeatletter
\hypersetup{pdftitle=\@title,pdfauthor=\@author}
\makeatother

\else

\usepackage{hyperref}
\hypersetup{colorlinks,bookmarksnumbered}

\makeatletter
\hypersetup{pdftitle=\@title,pdfauthor=\@author}
\makeatother

\fi 
\fi 

\def\va{{\vec a}}
\def\vk{{\vec k}}
\def\vr{{\vec r}}
\def\vq{{\vec q}}
\def\vp{{\vec p}}
\def\vb{{\vec b}}
\def\vd{{\vec d}}
\def\vl{{\vec l}}

\def\vK{{\vec K}}
\def\vG{{\vec G}}

\def\vA{{\vec A}}

\def\vB{{\vec B}}

\def\vp{{\vec p}}

\def\ve{{\vec e}}

\def\ve{{\vec e}}

\font\elevenmib=cmmib10 scaled 1095
\font\tenmib=cmmib10
\font\eightmib=cmmib10 scaled 800
\font\sixmib=cmmib10 scaled 667
\skewchar\elevenmib='177
\newfam\mibfam

\textfont\mibfam=\tenmib
\scriptfont\mibfam=\eightmib
\scriptscriptfont\mibfam=\sixmib

\mathchardef\sigma="711B

\def\ve{{\vec e}}

\def\dhat{{\hat {\vec d}}}

\def\ve{{\vec e}}

\newcommand{\be}{\begin{equation}}
\newcommand{\ee}{\end{equation}}
\newcommand{\bea}{\begin{eqnarray}}
\newcommand{\eea}{\end{eqnarray}}

\newcommand{\s}{\sigma}

\renewcommand{\phi}{\varphi}
\renewcommand{\epsilon}{\varepsilon}
\renewcommand{\vec}[1]{{\bf #1}}

\renewcommand{\Re}{{\mathrm{Re}}\, }
\renewcommand{\Im}{{\mathrm{Im}}\, }
\newcommand{\beq}{\begin{eqnarray}}
\newcommand{\eeq}{\end{eqnarray}}

\newcommand{\barr}{\begin{eqnarray}}
\newcommand{\earr}{\end{eqnarray}}

\mathchardef\sigma="711B

\def\ve{{\vec e}}

\def\ve{{\vec e}}

\mathchardef\sigma="711B

\def\vd{{\vec d}}
\def\ve{{\vec e}}

\def\ve{{\vec e}}

\ifodd 0
\renewcommand{\maketitlepage}{}
\renewcommand*{\makecopyrightpage}{}
\renewcommand*{\makeabstract}{}

\else

\abstract{
Graphene, the atomic-thin layer of carbon atoms, was first isolated on an insulating substrate in 2004 by two groups in Manchester  University \cite{Novoselov:2004,Novoselov:2005} and Columbia \cite{Zhang:2005}. Those milestone experiments established the Dirac nature of the charge carriers in graphene. The same year, C.L. Kane and E.G. Mele predicted that intrinsic spin-orbit coupling in graphene, if strong enough, would lead to a novel state of electronic matter called the Quantum Spin Hall (QSH) state \cite{Kane:2005a,Kane:2005b}. The QSH state is characterized by conducting gapless edge states circulating around an insulating bulk. Those edge states are protected from moderate disorder and interactions by a new topological invariant of the Z$_2$ nature. While the strength of spin-orbit coupling is too weak in graphene, it was soon predicted \cite{Bernevig:2006} and verified by transport experiments \cite{Konig:2007,Roth:2009} that the QSH state is realized in HgTe/CdTe quantum wells.  

In this manuscript, I will summarize some selected aspects of this huge field of research focused on Dirac matter including graphene and topological insulators. By Dirac matter, we have in mind various systems whose excitations obey a relativistic Dirac-like equation instead of the non relativistic Schrodinger equation. This report is mainly focused on the 2D topological insulators using graphene as a guideline. 

In chapter \ref{chapter0}, the semimetallic character of graphene is derived and the symmetry protection of the Dirac points are discussed while chapters \ref{chapter1} and \ref{chapter2} are devoted to Chern insulators and QSH insulators respectively.

}

\fi  

\begin{document}

\makefrontmatter



\chapter{Dirac fermions in graphene \label{chapter0}}

Graphene, the atomic-thin layer of carbon atoms, was first isolated on an insulating substrate in 2004 by two groups in Manchester  University \cite{Novoselov:2004,Novoselov:2005} and Columbia \cite{Zhang:2005}. Before those milestone experiments, it was already predicted that graphene should have a remarkable band structure \cite{Wallace:1947} hosting Dirac fermions as low energy excitations \cite{DiVincenzo:1984,Semenoff:1984,Haldane:1988}. Nevertheless it was commonly believed that a strictly (2D) two-dimensional carbon layer should also be unstable towards buckling or melting due to thermal fluctuations. Those experiments evidenced the existence of 2D Dirac fermions by the measurement of a very particular Quantum Hall effect \cite{Novoselov:2005,Zhang:2005} which is specific to relativistic carriers. It was also demonstrated that the density of such carriers can be tuned using a remote electrostatic gate, thereby realizing the first graphene-based field effect transistors \cite{Novoselov:2004}. Field effect transistors have a strong potential for applications in electronic devices, but they are also ideal systems to investigate the scattering properties of Dirac particles, including Klein tunneling \cite{Huard:2007,Williams:2007,Ozylmaz:2007,Huard:2009}.  

Here we would like to emphasize that the massless Dirac fermions in graphene are robust in many ways. Firstly they emerge at the level of non-interacting system, in contrast to other proposals of Dirac fermions at the nodes of $d-$wave superconductors \cite{Vishwanath:2001} or organic compounds \cite{Goerbig:2008}. Secondly, their zero mass character (gapless spectrum) is protected by the combination of time-reversal and inversion symmetries. Thirdly, interactions are quite inefficient in opening a gap or disrupting those quasiparticles \cite{CastroRMP:2009,GoerbigRMP:2011,KotovRMP:2012}. Finally spin-orbit coupling is also too weak to open a sizeable gap (see chapter \ref{chapter2}). 

In spite of this robustness, it is interesting, for both practical and fundamental purposes, to study how the Dirac points can be gapped out. In view of transistor applications, it is indeed mandatory to design narrow conducting channels by confining the carriers. Furthermore such conduction channels must be easily switchable between conducting (ON) and insulating (OFF) states. Hence it is crucial to understand and control how to turn semimetallic graphene into an insulator. On the fundamental side, those issues are closely related to the problem of mass generation for relativistic Dirac quasiparticles. Because they carry an internal isospin degree of freedom, Dirac quasiparticles can gain a mass in different ways. The various masses are characterized by matrices in isospin space and can be classified according to the symmetries they break or not. Increasing the number of internal degrees of freedom (that couple with the orbital motion) tends to create additional classes of insulators with very contrasted properties \cite{Ryu:2009}.

The physics of Dirac fermions in graphene has been extensively reviewed in far more details elsewhere \cite{CastroRMP:2009,GoerbigRMP:2011,KotovRMP:2012}. The aim of this short chapter is simply to emphasize the concept of mass (or gap) generation in graphene \cite{Ryu:2009} and to show that various insulating phases can be built (at least theoretically) from semimetallic graphene by adding proper perturbations. 

This chapter is organized as follows. We first show briefly how massless Dirac fermions appear as low-energy excitations of graphene. We then describe various microscopic models leading to finite gaps at the Dirac points and discuss the symmetries of such external perturbations. We restrict this chapter to spinless Dirac fermions where only four types of masses are possible. This restriction is relevant either when the spin is frozen (by an in-plane magnetic field for instance) or when the spin is totally decoupled from the orbital motion (which is the case in real graphene owing to the weakness of the spin-orbit interaction \cite{HuertasPRB:2006,MinPRB:2006}). More involved mass terms can appear when the spin-orbit coupling is included (chapter \ref{chapter2}) or when graphene is exposed to circularly polarized light.

\section{Massless Dirac fermions}

Here we briefly review the simplest tight-binding model for graphene (including only nearest-neighbor hopping amplitudes), and derive the corresponding low-energy theory. We discuss the Dirac nature of the low-energy theory and the protection of the Dirac points by fundamental symmetries. At neutrality (when undoped), graphene is a 2D semimetal with two isolated Fermi points. Upon raising/lowering the Fermi level (by adding/removing electrons), a 2D metal with electron-like/hole-like carriers is generated, and a circular Fermi surface is formed. The Dirac nature of such carriers is revealed by their scattering properties at scalar potential steps, in particular at an ambipolar ($pn$) junction \cite{Cheianov:2006,Katsnelson:2006,Cayssol:2009}. This Klein tunneling has been reported experimentally \cite{Huard:2007,Williams:2007,Ozylmaz:2007,Huard:2009} thereby providing a complementary proof of the Dirac nature of carrier beside the Quantum Hall measurements \cite{Novoselov:2005,Zhang:2005}.   

\subsection{Tight-binding model of graphene}
\begin{figure}
\begin{center}
\includegraphics[width=9cm]{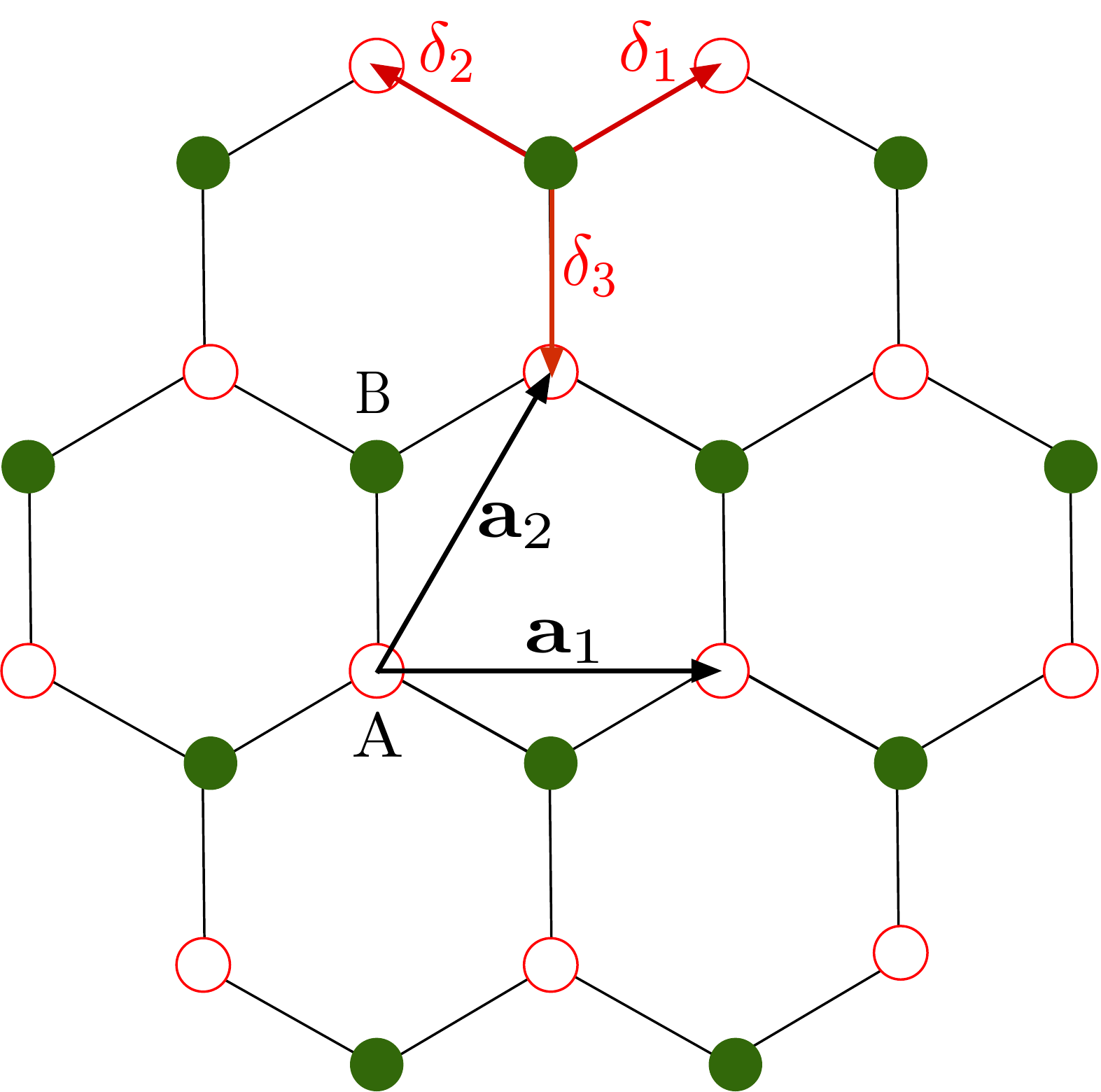}\\
\caption{Tight-binding model for graphene. Red open (green filled) dots for A (B) sublattice. Orange thick arrows for $\boldsymbol{\delta}_\alpha$ ($\alpha=1,2,3$). The distance between two sites is $a=0.142$ nm and the surface of the unit cell is $A_{cell}=3 \sqrt{3} a^2 /2$.}
\end{center}
\label{FigHoneycomb}
\end{figure}

{\it Honeycomb lattice.} Graphene consists of a honeycomb lattice of carbon atoms with two interpenetrating triangular sublattices, respectively denoted A and B (Fig. \ref{FigHoneycomb}). In this structure, each carbon atom has six electrons: two electrons filling the inner shell $1s$, three electrons engaged in the 3 in-plane covalent bonds $sp^2$, and a single electron occupying the $p_z$ orbital perpendicular to the plane. We are interested in the (2D )two-dimensional fluid formed by the $p_z$ electrons.   
Much of the physics of this 2D electronic system is described by the single orbital tight-binding Hamiltonian:
\begin{equation}
\label{GrapheneHamiltonian}
H_0 =  t \sum_{\langle i,j \rangle}    c^\dagger(\vr_i) c(\vr_j) =  t \sum_{\vr_A} \sum_{\alpha=1,2,3}   c_B^\dagger(\vr_A+\boldsymbol{\delta}_\alpha) c_A(\vr_A) + {\rm H.c.},
\end{equation}
where the sum $\langle i,j \rangle$ runs over nearest-neighbor (NN) sites $\vr_i$ and $\vr_j$, and $t \simeq -2.7$ eV is the hopping amplitude between the $p_z$ orbitals of two adjacent carbon atoms. The operator $c(\vr_i)$ destroys a fermion in the orbital $p_z$ at site $\vr_i$, and is also denoted $c_A (\vr_i)$ or $c_B (\vr_i)$ depending whether this site belongs to A sublattice or B sublattice. The sum over $\vr_A$ runs over the A-sites which form a triangular Bravais lattice spanned by the basis vectors:
\begin{equation}
\va_{1}=   \sqrt{3}  a \, \vec{e}_x ,     \,   \va_{2}=  \frac{a}{2} \left(    \sqrt{3} \vec{e}_x + 3 \vec{e}_y \right)    ,
\end{equation}
where $a=0.142$ {\rm nm} is the length of the carbon-carbon bond. The vectors $\boldsymbol{\delta}_{\alpha}$ defined by 
\begin{equation}
\boldsymbol{\delta}_{1,2}=   \frac{a}{2} \left(   \pm \sqrt{3} \vec{e}_x +  \vec{e}_y \right)   ,     \,   \boldsymbol{\delta}_{3}=   -  a \, \vec{e}_y ,
\end{equation}
connect any A-site to its three B-type nearest neighbors (Fig. \ref{FigHoneycomb}). The hopping matrix elements between next-nearest neighbors (and more distant atoms) are neglected which is justified by the fact that those corrections are roughly ten times smaller than the main hopping $t$. 

\bigskip
  
{\it Electronic band structure.} Owing to translation invariance, the two-dimensional momentum $\vk$ is a good quantum number. In order to diagonalize the Hamiltonian Eq. (\ref{GrapheneHamiltonian}), we use the Fourier transformation:  
\begin{equation}
\label{Fourier}
c_a(\vr_i)= \frac{1}{\sqrt{N}} \sum_{\vk} e^{-i \vk . \vr_i} c_{a}(\vk), 
\end{equation}
where $a=A,B$ is the sublattice index and $N$ is the total number of sites. After substitution of Eq.(\ref{Fourier}), the Hamiltonian Eq. (\ref{GrapheneHamiltonian}) becomes diagonal in momentum:
\begin{equation}
\label{AGrapheneHamiltonianFourier}
H_0  = t \sum_\vk \gamma(\vk)    c_B^\dagger(\vk) c_A(\vk) + {\rm H.c.},
\end{equation}
where one has defined the quantity: 
\begin{equation}
\label{FourierGamma}
\gamma(\vk) = \sum_{\alpha=1,2,3} e^{i \vk . \boldsymbol{\delta}_\alpha}=2e^{ik_y a/2} \cos \frac{\sqrt{3} k_x a}{2} + e^{-i k_y a} .
\end{equation}
The electronic energy spectrum is given by 
\begin{equation}
\label{SpectrumTBM}
E(\vk)=\pm t |\gamma(\vk)|,
\end{equation}
which describes a valence band (minus sign) and a conduction band (plus sign) that are symmetric with respect to $E=0$. Note that this zero of energy corresponds to the common energy of the atomic orbitals on sublattices $A$ and $B$. The valence and conduction bands touch at isolated points of the first Brillouin zone (FBZ) obtained by solving  the equation $\gamma(\vk)=0$. Mathematically those touching points, also called Dirac points (for a reason explained in the next paragraph), span an infinite array of discrete locations of the reciprocal space. Of course there is a huge redundancy and all the points linked by a vector of the reciprocal lattice actually describe the same physical state. There are only two inequivalent Dirac points:
\begin{equation}
\vk = \pm \vK=\pm \frac{4 \pi}{3\sqrt{3}a} \, \vec{e}_x ,
\end{equation}
in the FBZ. Other solutions of the equation  $\gamma(\vk)=0$ can be linked by a reciprocal lattice vector to one of those two solutions, and therefore describe the same physical state. 

Note that the existence of isolated solutions of  $\gamma(\vk)=0$, preventing the system to become gapped, is robust even if some crystal symmetries are lost and more hopping amplitudes are added. For instance additional second-neighbor hoppings will break the electron/hole symmetry discussed above, but will not affect the existence of Dirac points. Other perturbations, like a uniform anisotropic deformation on one type of bond, only shift the Dirac points and modify the conical dispersion around them \cite{Goerbig:2008}. In fact the touching points are protected by more fundamental symmetries, namely inversion and time-reversal symmetries as we will show in detail later (\ref{microsymmetries}).

\bigskip

{\it Electronic filling.} We now discuss how the electronic states are filled by electrons. Since the system has one electron per orbital (or site), the bands described above would be completely filled (both valence and conduction) if were ignoring spin. Due to the twofold spin degeneracy of each $\vk$-state, neutral graphene corresponds exactly to the situation of half-filling of the $\pi$ bands, and the Fermi level is at $E=0$. Therefore the valence band is completely filled at zero temperature while the conduction band is completely empty, the Fermi level being reduced to the two contact points at $\vk = \pm \vK$. The total density of $p_z$ electrons, $n_{neutrality}=2/A_{cell}\simeq 4.10^{15}$ cm $^{-2}$ calculated as 2 electrons per unit cell (1 electron per carbon atom) is just the density necessary to fill completely the valence band. This density is not to be confused with the density of carriers $n=n_{total}-n_{neutrality}$ which is zero at neutrality and can be positive or negative depending whether electrons are actually added or removed from the crystal. The carrier density is usually smaller than $n_{neutrality}$, typically $n \simeq 10^{11}-10^{13}$ cm$^{-2}$ otherwise the Dirac Hamiltonian is no longer a good approximation for larger level of dopings (and for the corresponding high energies).   

\begin{figure}
\begin{center}
\includegraphics[width=9.5cm]{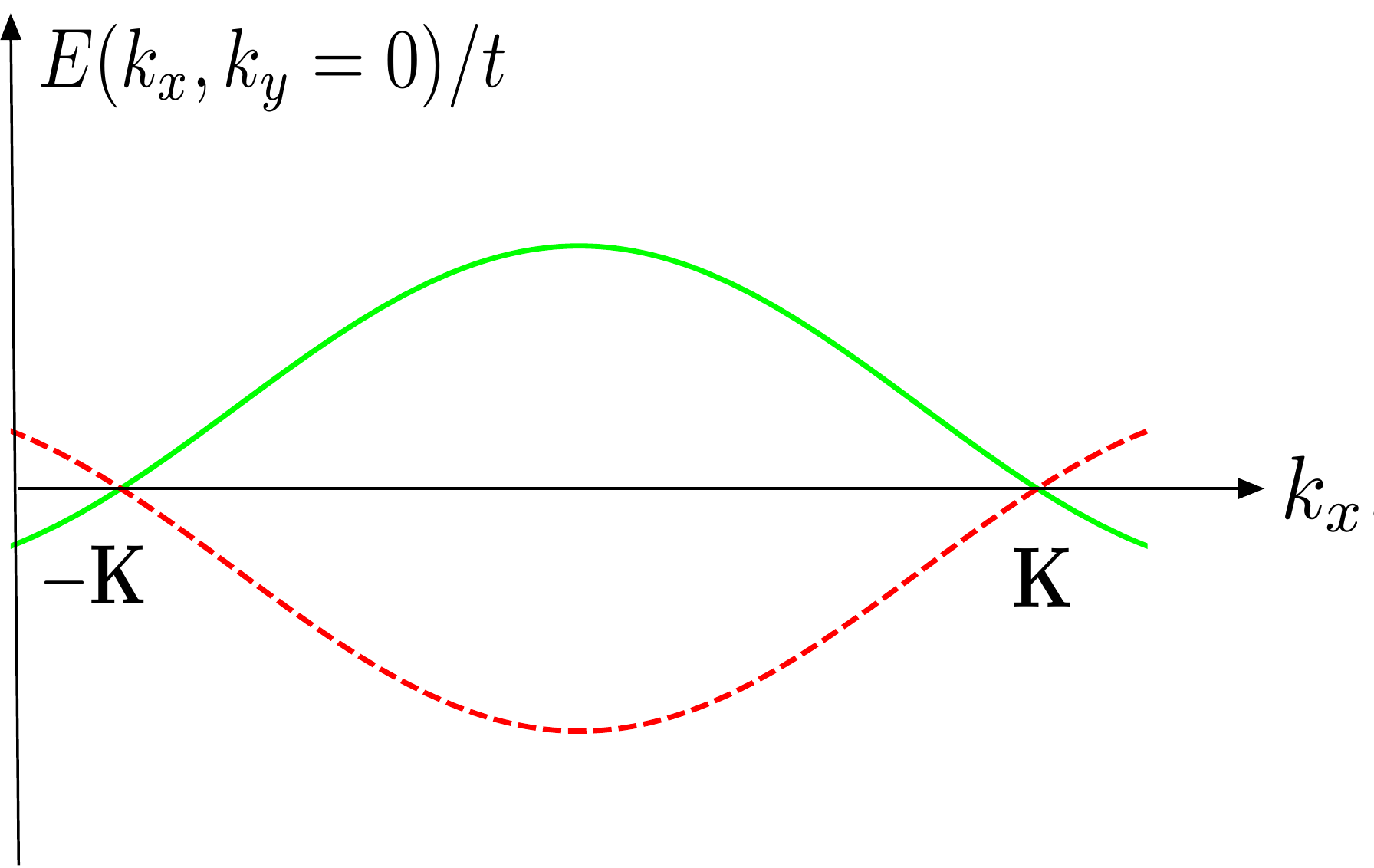}
\caption{Electronic energy dispersion $E(\vk)=\pm t |\gamma(\vk)|$ of graphene for $k_y=0$.}
\end{center}
\label{FigBandsHighEnergy}
\end{figure}

\subsection{Low energy theory near the Dirac points}

{\it Dirac Hamiltonian.} We consider now the low-energy theory for the single-particle states near the Dirac points. The momenta are written as $\vk =\pm  \vK + \vq$ close to the zero energy points ($|\vq|a \ll 1$), and the annihilation operators for those states are denoted $c_{A\pm\vK} (\vq)=c_{A}(\pm\vK+\vq)$ where $\vq = q_x \ve_x +  q_y \ve_y$ is a small momentum deviation from the Dirac points. From the first-order expansion of $\gamma(\vk)$ around the Dirac points:
\begin{equation}
\label{lineargamma}
\gamma(\pm \vK +\vq)=-3a(\pm q_x + i q_y)/2 ,
\end{equation}
we find that the Hamiltonian describing the low energy excitations near $\vk=\xi \vK$ ($\xi=\pm 1$) can be approximated as:
\begin{equation}
H_0^{(\xi \vK)} =v_F \sum_\vk
\begin{pmatrix} 
c_{A\xi\vK}^\dagger(\vq) & c_{B\xi\vK}^\dagger (\vq)   \\
\end{pmatrix}
\begin{pmatrix} 
0 & \xi q_x -i q_y \\
 \xi q_x  + i q_y & 0 
\end{pmatrix}
\begin{pmatrix} 
c_{A\xi\vK} ( \vq) \\
c_{B\xi\vK} (\vq) 
\end{pmatrix},
\end{equation}
where $v_F = -3 at/2 \simeq 10^6$ m.s$^{-1} \simeq c/300$ is the Fermi velocity. The Fermi velocity is basically the bandwidth $t$ divided by the Brillouin zone (BZ) size $1/a$. It is convenient to introduce a spinor representation:
\begin{equation} 
c_a^\dagger(\vq)=( c_{A\vK}^\dagger  c_{B\vK}^\dagger   c_{A-\vK}^\dagger  c_{B-\vK}^\dagger  ),
\end{equation} 
embedding the two zero energy points $\xi=\pm 1$. Then the single-electron Hamiltonian can be written in the compact form:
\begin{equation}
H_0 = \sum_{\vq} \sum_{a,b=1}^{4} c_a^\dagger(\vq) (\mathcal{H}_{0}(\vq))_{ab}c_b(\vq) ,
\label{Hamiltonianzerospinor}
\end{equation}
with:
\begin{equation}
\mathcal{H}_{0}(\vq) =v_F  (q_x  \sigma_x \tau_z+ q_y  \sigma_y) ,
\label{BlochHamiltoniangraph}
\end{equation}
which has exactly the form of the Dirac Hamiltonian describing a spin one-half relativistic particles with zero mass. In particular the dispersion relation is simply:
\begin{equation}
E(\vq) =v_F  |\vq| ,
\label{DispersionMasslessDirac}
\end{equation}
typical of a relativistic massless particle with velocity of light replaced by $v_F$.

Nevertheless we would like to emphasize the differences between the Dirac equation in the context of graphene system and in the high-energy framework respectively. In high-energy physics, the Dirac equation comes from Lorentz-invariance and very general considerations to associate special relativity and quantum mechanics. Then the minimal objet to satisfy such an equation is a bispinor combining the spin and particle/hole symmetry relating positive and negative solutions of the Dirac equation. The coherent interpretation of the negative energy states led Dirac to the prediction of antiparticles which led to the discovery of the positron.

In graphene, the origin of the Dirac physics is totally different. As we have seen, the spinors originate from a $\vk .\vp$ expansion around special points of a particular band structure. Hence in graphene, there is no fundamental issue with the negative energy states that are just the valence band states (these states are in fact bounded from below by the bottom of the valence band). Finally the emergent Lorentz invariance of Eq. (\ref{DispersionMasslessDirac}) is only valid near the Dirac point, namely for wave vectors $\vq$ located in a disk whose radius is far smaller than the inverse lattice spacing $1/a$, whereas Lorentz invariance applies in the whole Minkowski space-time in particle physics.

\bigskip

{\it How many Dirac cones ?} Fundamentally graphene is a two-band system because it has 2 orbitals per unit cell (one $p_z$ orbital per atom and 2 atoms in the unit cell), thereby having two states per momentum $\vk$ in the first Brillouin zone (FBZ). Due to fermion doubling, there are two species of massless Dirac fermions (one for each valley) carrying a sublattice isospin coupled to their momentum. Note that the presence of 4$\times$4 matrices in Eq.(\ref{BlochHamiltoniangraph}) does not mean that graphene is a 4 band system in the same sense as the genuine 4 band insulators we shall study in chapter \ref{chapter2}. Indeed for a given value of $\vk$ in the FBZ graphene has only two states (Fig. \ref{FigBandsHighEnergy}). Besides, when discussing transport or at least ballistic elastic scattering at the Fermi level, there are effectively 4 states sharing the same energy (Fig. \ref{FigBandsHighEnergy}). Then the physics depends on the ratio between intravalley and intervalley scattering rates. For instance, Klein tunneling and weak antilocalization are better observed if the intervalley coupling is much weaker than the intravalley coupling.  

Here we have not considered explicitly the real spin simply because it is not coupled to the momentum in the absence of spin-orbit. In fact at each valley, there are two completely degenerated and decoupled Dirac cones corresponding to each spin direction. Hence graphene has 4 Dirac cones with sublattice-momentum locking. This is at odds with surface state of 3D strong topological insulators which has a single Dirac cone with momentum coupled to the real spin. 
 
 \bigskip

\subsection{Microscopic symmetries \label{microsymmetries}}

In graphene, the fundamental symmetries are time-reversal symmetry $\mathcal T$ and inversion symmetry $\mathcal P$. Again we restrict our discussion to spinless fermions. 

\bigskip

{\it Time-reversal symmetry.} In order to see how time-reversal symmetry acts in the effective low-energy theory, we first consider the wave function in the microscopic theory:
\begin{equation}
\Psi(\vr_A)=\phi_{A+}(\vr) e^{i\vK . \vr_A}+\phi_{A-}(\vr) e^{-i\vK . \vr_A}.
\end{equation}
The time-reversed state is described by the complex conjugated wave function, namely:
\begin{equation}
\Psi^{*}(\vr_A)=\phi_{A+}^{*}(\vr) e^{-i\vK . \vr_A}+\phi_{A-}^{*}(\vr) e^{i\vK . \vr_A}.
\end{equation}
Now each Fourier mode $u_{A\pm}(\vq) e^{i \vq . \vr}$ of the envelope functions $\phi_{A\pm}(\vr)$ is transformed into $u_{A\mp}(\vq)^{*} e^{-i \vq . \vr}$. Of course, one can proceed similarly for any site on the B sublattice. Hence in momentum representation, the effect of time-reversal transformation is to complex conjugate the amplitudes, change $\vq$ into $-\vq$, and interchanges the valley labels while leaving unchanged the sublattice index. In the representation $c_{a}^\dagger(\vq)=( c_{A\vK}^\dagger  c_{B\vK}^\dagger   c_{A-\vK}^\dagger  c_{B-\vK}^\dagger  )$, the operations described above can be summarized as:
\begin{equation}
\mathcal T =\tau_x K_c,
\end{equation}
where $K_c$ complex conjugate the amplitudes of the spinor and changes $\vq$ into $-\vq$ in momentum representation. It is very important to note that this time-reversal operation obeys $\mathcal{T}^2=1$ because we deal with spinless fermions. When the spin is included (see chapter \ref{chapter2}), the full time-reversal operation square to $-1$ with Kramers degeneracy as a fundamental consequence.  

\bigskip

{\it Inversion symmetry.} The inversion symmetry switches the sublattice $A$ and $B$, and also changes the momentum $\vk$ into $-\vk$. This means in particular that the valleys are switched and that the mode $e^{i \vq \vr}$ of the envelope function in valley $\xi$ becomes the mode $e^{-i \vq \vr}$ in the opposite valley $-\xi$. Hence the inversion operation $\mathcal P$ can be written as:
\begin{equation}
\mathcal P =\sigma_x \tau_x ,
\end{equation}
and $\vq  \rightarrow  -\vq$. 

{\it Low energy theory}
Now we can check that the low-energy 4$\times$4 Bloch Hamiltonian Eq.(\ref{BlochHamiltoniangraph}) is invariant both by the time-reversal operation $\mathcal T$:
\begin{equation}
\mathcal{T} \mathcal{H}_{0}(\vq) \mathcal{T}^{-1}= v_F \tau_x \left(   q_x  \sigma_x \tau_z+ q_y  \sigma_y^*  \right) \tau_x = \mathcal{H}_{0}(-\vq),
\end{equation}
and by the inversion operation:
\begin{equation}
\mathcal{P} \mathcal{H}_{0}(\vq) \mathcal{P}^{-1}= v_F \sigma_x \tau_x \left(   q_x  \sigma_x \tau_z+ q_y  \sigma_y  \right) \sigma_x \tau_x  =\mathcal{H}_{0}(-\vq) ,
\end{equation}
where we have used the algebra of Pauli matrices. Using those symmetry operators, we can check the symmetry of any perturbation (or mass term) written in this low energy sector.

\section{Massive Dirac fermions}

We have seen that the combination of $\mathcal T$ and $\mathcal P$ protect the gapless nature of spinless fermionic excitations in graphene. In 1984, Semenov discussed the effect of a sublattice staggered potential, which breaks the equivalence between A and B sites, thereby generating a gap at the Dirac points   \cite{Semenoff:1984}. In a seminal paper \cite{Haldane:1988}, Haldane introduced a special periodic magnetic field which also leads to massive Dirac fermions and further provides a lattice realization of the Quantum Hall Effect in the absence of a net magnetic flux through the unit cell. Finally it was recently emphasized that a Kekule-type dimerization pattern of the carbon-carbon bonds can also open gaps at the Dirac points. In this report we call the corresponding insulating states the Semenov, the Haldane and the Kekule insulators respectively. In the low energy theory, a mass is a perturbation acting on the wave functions and able to open a gap. Such a perturbation is represented by a 4$\times$4 matrix that anticommutes with the kinetic/velocity part $\mathcal{H}_0(\vq)$ of the graphene Hamiltonian. 

\subsection{Semenov insulator} 

\begin{figure}
\begin{center}
\includegraphics[width=6.5cm]{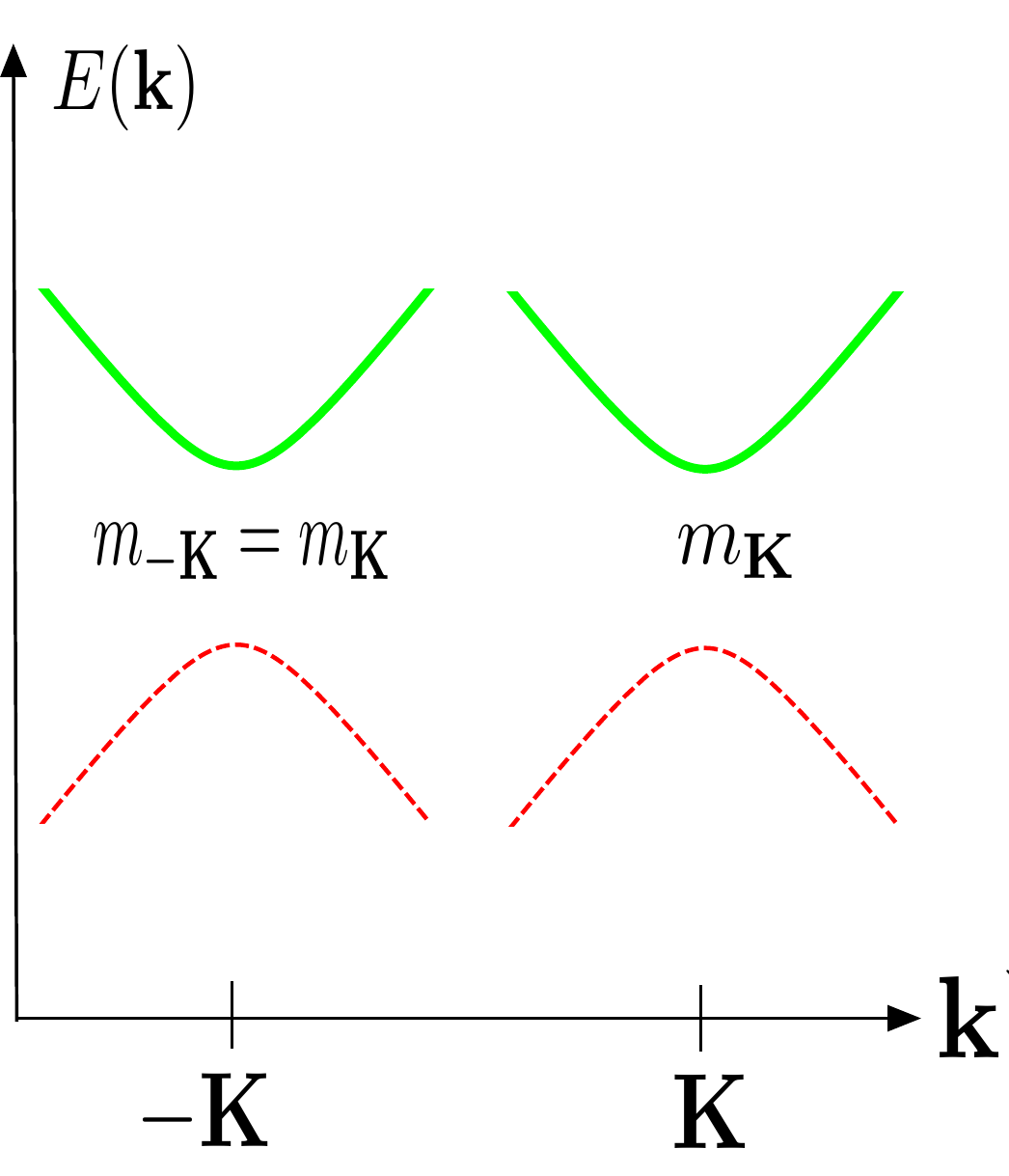}
\hspace{1cm}
\includegraphics[width=6.5cm]{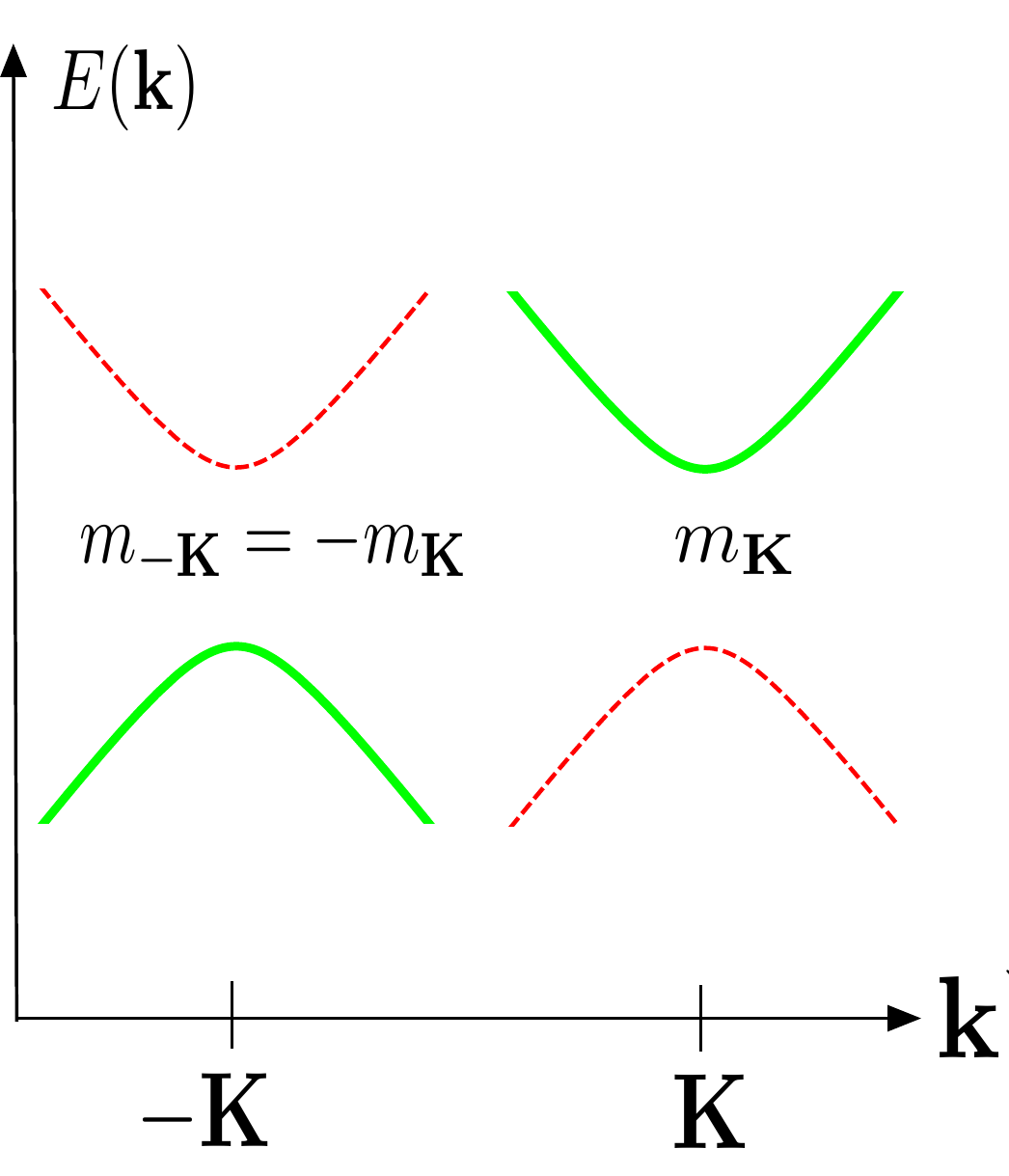}
\caption{Low energy dispersion for the Semenov insulator and Haldane insulator.}
\end{center}
\label{FigSemenovHaldaneBands}
\end{figure}
A local on-site staggered potential described by 
\begin{equation}
\label{TBSemenoff}
H_1  =   \sum_{r_A}  \epsilon_A c_A^\dagger(\vr_A) c_A(\vr_A) +  \sum_{r_B}  \epsilon_B c_B^\dagger(\vr_B) c_B(\vr_B) ,
\end{equation}
spoils the equivalence between orbital energies $\epsilon_A$ and $\epsilon_B$ on sites A and B respectively, and therefore also breaks the inversion symmetry $\mathcal{P}$. This situation is naturally realized for a honeycomb structure where the A and B sites are actually occupied by different atoms, like in BN crystals. Being local, this perturbation term is non dispersive ($\vk$-independent) and its Bloch Hamiltonian written is lattice isopin$\otimes$valley space reads
\begin{equation}
\mathcal{H}_{1} (\vk)= \mathcal{H}_{1}= \epsilon_0 {\bold I}+ M_1 \,  \sigma_z  ,
\label{SemenovLowTerm}
\end{equation}
which is typical of a massive relativistic particle. The mass is given by $M_1=(\epsilon_{A}-\epsilon_{B})/2$. The first term, proportional to identity $\bold I$, only shifts the position of the Fermi level without changing the dispersion of the states. The $\sigma_z = \sigma_z \otimes I_\tau$ term anticommutes with $\mathcal{H}_0(\vq)$ and therefore can open a gap of size $2|M_{1}|$ at both Dirac points. The spectrum, obtained by squaring the Hamiltonian $\mathcal{H}_0(\vq) + M_1 \sigma_z$ and using $\{\sigma_z ,\mathcal{H}_{0}\}=0$, is:
\begin{equation}
E(\vq)  = \pm \sqrt{v_F^2 q^2 + M_1^2} .
\label{SpectrumSemenov}
\end{equation}
There is a gap opening only if $\epsilon_A \neq \epsilon_B$, namely if A and B sites are non equivalents. We finally check explicitly that the perturbation $\mathcal{H}_1(\vq)$ is is odd under inversion $\mathcal{P}=\tau_x \sigma_x$
\begin{equation}
\mathcal P \mathcal{H}_1 \mathcal {P}^{-1}=- \mathcal{H}_1,
\end{equation}
and even under time-reversal. 

\subsection{Haldane insulator}

Haldane introduced a model of graphene under a modulated magnetic field that fully respects the spatial symmetries of the Bravais lattice but breaks $\mathcal{T}$. This model was motivated by realizing the Quantum Hall effect without a global magnetic field by unit cell and without the Landau level structure. To realize that one should break time-reversal symmetry by inserting local fluxes which sum up to zero over each unit cell. This field preserves translational symmetry and the Bloch nature of electronic states. Those fluxes can be described by introducing unimodular phase factors in the second neighbor hopping amplitudes $t_2\rightarrow t_2 e^{\pm i \phi}$, where the $\pm$ sign corresponds to the different chiralities. 

The microscopic Haldane model is given by the Hamiltonian $H=H_0 + H_1 + H_2$. The first perturbation $H_1$ is the staggered on-site potential, Eq. (\ref{TBSemenoff}), first considered by Semenov and discussed in the previous paragraph. Let us now focus here on the additional second-neighbor hopping term introduced by Haldane: 
\begin{equation}
\label{HaldaneHamiltonian}
H_{2}  =   t_2 \sum_{\langle \langle i,j \rangle \rangle} e^{i \nu_{ij} \phi}  c^\dagger_i c_j, 
\end{equation}
where the sum runs over the next-nearest neighbor (NNN) sites $\langle \langle i,j \rangle \rangle$. The chirality $\nu_{ij}= \pm 1$ is defined as follows. Let us consider two NNN sites denoted $i$ and $j$. We call $\dhat_{ij}^{(1)}$ the unit vector from site $i$ to the intermediate site (linking $i$ and $j$), and $\dhat_{ij}^{(2)}$ the unit vector from this intermediate site to $j$. Then 
\begin{equation}
\label{chirality}
\nu_{i,j} = \left( \dhat_{ij}^{(1)} \wedge \dhat_{ij}^{(2)} \right).\vec{e}_z .
\end{equation}
To be more specific and to write explicitely the sublattice structure of those Haldane hopping terms, let us introduce the vectors $\vb_1=\boldsymbol{\delta}_{2}-\boldsymbol{\delta}_{3}$, $\vb_2=\boldsymbol{\delta}_{3}-\boldsymbol{\delta}_{1}$, and $\vb_1=\boldsymbol{\delta}_{2}-\boldsymbol{\delta}_{3}$, connecting next-nearest neighbor sites. Then the Hamiltonian $H_2$ reads:
\begin{equation}
\label{GrapheneH2}
H_2 =  t_2   \sum_{i}   \left( \sum_{\vr_A} c_A^\dagger(\vr_A) c_A(\vr_A +\vb_i) e^{i \phi} +\sum_{\vr_B}   c_B^\dagger(\vr_B) c_B(\vr_B + \vb_i) e^{-i\phi} \right) + {\rm H.c.},
\end{equation}
where $i=1,2,3$.
After Fourier transform, this Hamiltonian becomes 
\begin{equation}
\label{HaldaneHamiltonianFourier}
\mathcal{H}_{2}(\vk)  =  2 t_2  \left[ \cos(\phi)   \sum_{i=1,2,3}       \cos(\vk . \vb_i) {\bold I}  +   \sin(\phi)   \sum_{i=1,2,3}   \sin(\vk . \vb_i)  \sigma_z \right]   ,
\end{equation}
which is valid for any $\vk$ in the FBZ.

\begin{figure}
\begin{center}
\includegraphics[width=7cm]{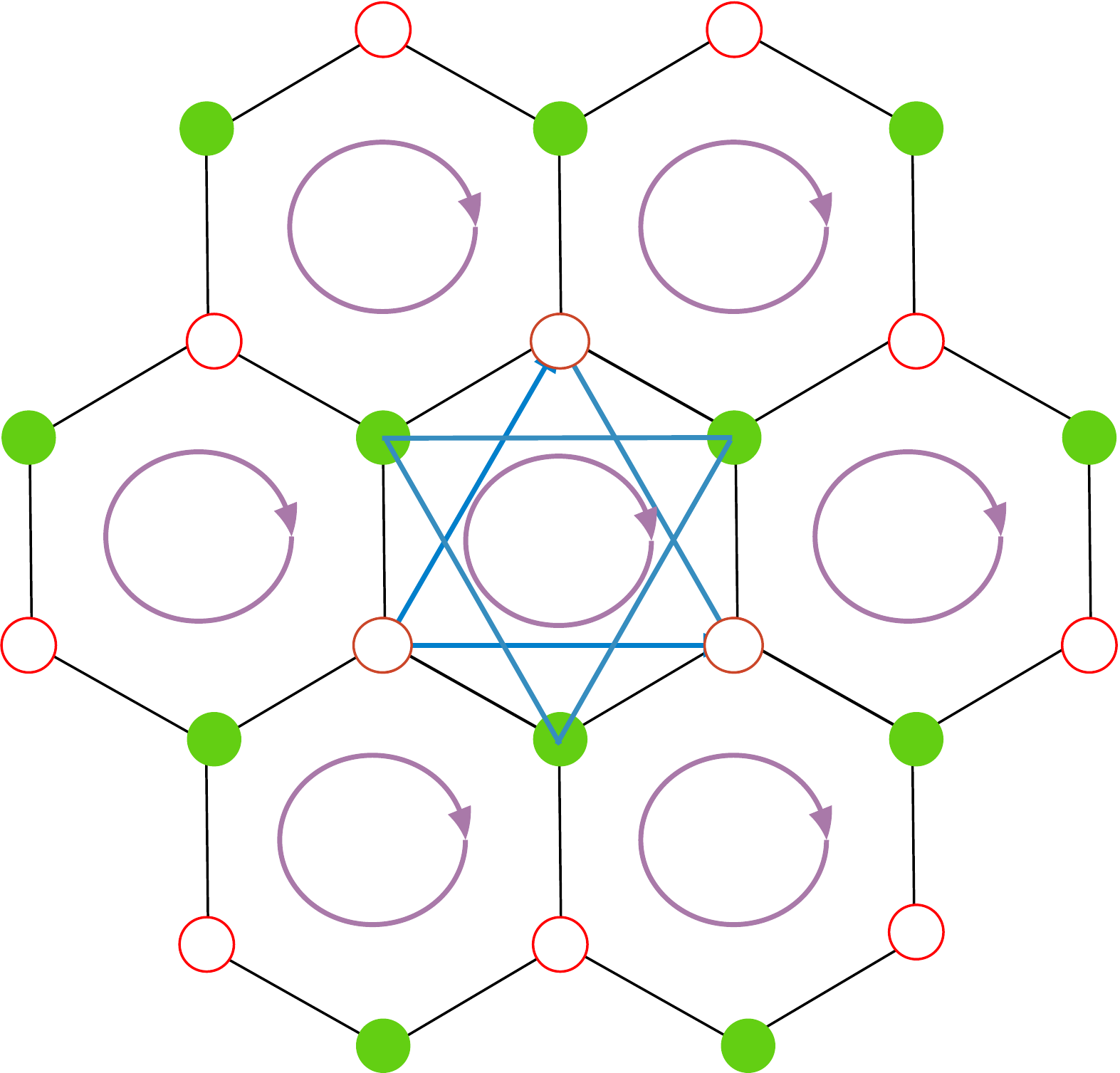}\\
\caption{Tight-binding model for the Haldane model. Blue arrows (direction indicated by the oriented loops inside the hexagons) stands for $t_2 e^{i \phi}$. The vectors connecting NNN neighbor sites are defined as $\vb_1=\boldsymbol{\delta}_{2}-\boldsymbol{\delta}_{3}$, $\vb_2=\boldsymbol{\delta}_{3}-\boldsymbol{\delta}_{1}$, and $\vb_3=\boldsymbol{\delta}_{1}-\boldsymbol{\delta}_{2}$.}
\end{center}
\label{FigHaldaneModel}
\end{figure}

The $NNN$ perturbation is dispersive ($\vk-$dependent) because the it is nonlocal in real space. Near the Dirac points, one has simply to substitute $\vk=\vK$ (or $\vk=-\vK$) as a zero order approximation. Here we focus on states in each valley and obtain:
\begin{equation}
\label{Sums}
\sum_{i=1,2,3}   \cos(\vK . \vb_i)   =  - \frac{3}{2} ,   \,    \,    \,        \,        \,
\sum_{i=1,2,3}  \sin(\pm \vK . \vb_i) = \mp \frac{3 \sqrt{3}}{2} ,
\end{equation}
where we have used that $\vK.\vb_1=\vK.\vb_2=\vK.\vb_3=4\pi/3$. The part of $\mathcal{H}_2$ which is proportional to the identity just shifts the energies and spoils the electron-hole symmetry of the purely NN model. Nevertheless the system remains gapless (if $\phi=0,\pi$) under introduction of a real NNN hoppings. This is because both $\mathcal{P}$ and $\mathcal{T}$ are preserved for real NNN hoppings. In contrast for complex hoppings, the term proportional to $\sigma_z $, denoted $\mathcal{H}^{\rm trb}_2(\vk)$ hereafter, opens a gap at the Dirac points. In the low energy model, this Haldane mass term is \cite{Haldane:1988}:
\begin{equation}
\mathcal{H}^{\rm trb}_{2}(\pm \vK)  = -3 \sqrt{3} t_2 \sin (\phi) \,  \sigma_z \tau_z  ,
\label{HaldaneTerm}
\end{equation}
which changes sign in different valleys. This is at odds with the Semenov insulator where both valleys are characterized by the same gap. The spectrum is again obtained by squaring the Bloch Hamiltonian $\mathcal{H}^{\rm trb}_{2}(\vq)$ and using the anti commutation property of the Pauli matrices:
\begin{equation}
E  =\pm \sqrt{v_F^2 p^2 + 27 t^2 \sin^2\phi}.
\label{Spectrum}
\end{equation}

This Bloch Hamiltonian breaks time-reversal symmetry because it is odd under the operator $\mathcal T$:
\begin{equation}
\mathcal T \mathcal{H}^{\rm trb}_2(\vq=0) \mathcal {T}^{-1}=- \mathcal{H}^{\rm trb}_2(\vq=0),
\end{equation}
but it is even under $\mathcal P =\tau_x \sigma_x$.
One can notice that this term cancels for $\phi=0$ where the time-reversal is trivially restored. This is also the case at $\phi=\pi$ because 
$e^{i\pi}=e^{-i\pi}=-1$.

\subsection{Kekule insulator} 
The real-valued modulation of the nearest-neighbor hopping amplitude can open a gap when the wave vector of the modulation is connecting the two valleys. The Hamiltonian is $H_0 + H_3$ where the texture in the hopping amplitudes is described by \cite{Ryu:2009}: 
\begin{equation}
\label{KekuleHamiltonian}
H_3 =  \sum_{\vr_A} \sum_{\alpha=1,2,3} \delta t_\alpha(r_A)  c_B^\dagger(\vr_A+\boldsymbol{\delta}_\alpha) c_A(\vr_A) + H.c.,
\end{equation}
where:
\begin{equation}
\label{Kekule}
\delta t_\alpha(r_A) = \Delta(\vr_{A}) e^{i \vK \delta_\alpha} e^{i G.\vr_A}.
\end{equation}
The vector $\vG = 2 \vK $ connects the two Dirac points. The low-energy corresponds to the mass terms $\tau_{x,y} \sigma_z$.

The Semenov and the two Kekule masses belongs to the same triplet of the representation of Clifford algebra by 4$\times$4 matrices 
while the Haldane mass is a singlet. This already signals that Haldane mass has a particular role that we shall explore in the next chapter on Chern insulators.

\section*{Conclusion and perspectives}

In this chapter, we have briefly reviewed that graphene has two Dirac points protected by the simultaneous presence of time-reversal and inversion symmetries. Those Dirac points can be gapped out by perturbations that break one of those symmetries. 
We have discussed only spinless fermions in graphene but similar questions can be raised for any semimetal where the carrier motion is coupled to one (or several) internal degrees of freedom. 

The number of distinct mass terms increases with the number of internal degrees of freedom that couple with the electronic motion. For the case of spinless fermions, there are four types of masses, namely the inversion breaking (Semenov) mass, the time-reversal breaking (Haldane) and two Kekule masses. Interestingly those insulating phases have very contrasted properties. The Semenov insulator is similar to an ordinary band insulator characterized by a purely local response to electric fields. In contrast the Haldane insulator has a chiral edge mode that conduct electricity non locally around an insulating bulk. The Kekule insulator can host fractionalized excitations. Finally the inclusion of spin (and spin-orbit coupling) can lead to even more complex phases, including topological ones like the Quantum Spin Hall insulator (see chapter \ref{chapter2}). 

Finally, the general idea is that insulators are not all the same because they can differ by the topology of their Bloch wave functions even when their energy spectra are identical. This is at odd with usual non relativistic carriers in standard semiconducting systems where the gap is simple scalar and topologically protected surface states are absent.

\chapter{Chern insulators \label{chapter1}}

Chern insulators (CIs), also called Quantum Anomalous Hall (QAH) phases, are band insulators that exhibit the Quantum Hall effect (QHE) in the absence of an overall external magnetic field. Historically the first CI model was devised by Haldane \cite{Haldane:1988} who demonstrated that carefully chosen complex second neighbor hopping terms on the honeycomb lattice (chapter \ref{chapter0}) generate a QHE carried by Bloch bands (instead of Landau levels). More recently the advent of topological insulators  \cite{KaneRMP:2011,QiRMP:2011,KonigJPSJ:2008,QiPhysToday:2010} has renewed the interest on the Haldane model and initiated the investigation of many models of CIs on various lattices \cite{Qi:2006,Tang:2011,Sun:2011,Neupert:2011,Sticlet:2012}. The Haldane insulator is also interesting in the doped regime where it becomes a metal with Berry phase effects and in particular where it exhibits the Anomalous Hall effect \cite{AQHRMP}.  

In order to generate a nonzero charge Hall conductance, it is necessary to break time-reversal (TR) symmetry. In the standard QHE, time-reversal symmetry is broken by a strong magnetic field which also creates the Landau level structure. In CI models the mandatory symmetry breaking is accomplished by introducing a periodic pattern of local fluxes that respects the full translational symmetry of the crystal. This makes CIs interesting in many respects. First, CIs differ from ordinary (atomic or covalent) band insulators by the presence of conducting edge channels. Second, CIs offer the opportunity to revisit the Quantum Hall physics in lattice systems which a priori differ from the usual continuous two-dimensional electron gas experimentally relevant for Si, GaAlAs and graphene-based quantum Hall devices.

In practice generating Bloch bands with finite Chern number requires a very demanding band-structure engineering which is extremely difficult to achieve experimentally. In real materials, CI would require a nontrivial magnetic background, arising from a subtle interplay of spin-orbit (or some coupling between momentum and an internal degree of freedom) and exchange field effects (necessary to break TR symmetry). Another route could be to realize Chern (or QAH) phases under non equilibrium conditions \cite{Oka:2009,Kitagawa:2011,lindner,Cayssol:2013}.

This chapter is organized as follows. In Section \ref{sectionTBI} the general model of a two-band insulator with Berry phase effects in crystals is presented \cite{XiaoBerryRMP:2010}. The section \ref{sectionChern} is devoted to the concept of (Chern) topological invariant and its relation to the QHE in Chern insulators. Finally the mechanism for the formation of topologically protected edge states is explained by solving several simple models of an interface between two distinct gapped phases (section \ref{sectionEdge}). Conceptually CIs (or QAH phases) are the simplest topological phases, in terms of number of bands, that can be realized in a crystal: the simplest CIs are two-band insulators that provide building blocks for the understanding of more involved topological models like the QSH states (see chapter \ref{chapter2}).

 \section{Two-band insulators \label{sectionTBI}}
Here we introduce the general framework to study Berry phase and topological effects in crystals using the simplest possible system: the two-band insulator (in mostly 2 dimensions). Two-band insulators play a fundamental role for the understanding of the classification of electronic band structures, akin to the the two-level system in atomic physics.

\subsection{Bloch Hamiltonian}
Let us consider a $d$-dimensional lattice model for a two-band insulator. The tight-binding Hamiltonian in real space can be written as:
\begin{equation}
 H=\sum_{\vr_i,\vr_j} c_a^\dagger(\vr_i) h_{ab}(\vr_i - \vr_j) c_{b}(\vr_j),
\label{Hamiltonianbands}
\end{equation}
where the sum over internal index $a,b=1,2$ is implied. Translational invariance allows to use the $d$-dimensional momentum $\vk$ as a good quantum number. Like for graphene (Eq.\ref{Fourier}), one introduces the Fourier transform of the second-quantized operators:
\begin{equation}
\label{FourierChern}
c_a(\vr_i)= \frac{1}{\sqrt{N}} \sum_{\vk} e^{-i \vk . \vr_i} c_a(\vk),
\end{equation} 
where $N$ is the total number of lattice sites.
This allows to write the Hamiltonian as:
\begin{equation}
 H=\sum_{\vk} c_a^\dagger(\vk) \mathcal{H}_{ab}(\vk) c_{b}(\vk) ,
\label{HamiltonianbandsFourier}
\end{equation}
where $\mathcal{H}(\vk)$ is a $2\times 2$ Hermitian matrix defined by:
\begin{equation}
 \mathcal{H}_{ab}(\vk)=\sum_{\vr} h_{ab}(\vr) e^{i \vk . \vr}.
\label{HamiltonianbandsBlochFourier}
\end{equation}
This Bloch Hamiltonian is then a $2\times 2$ Hermitian matrix, acting on the Bloch spinors, and that can be parametrized as
\begin{equation}
\mathcal{H}(\vk)=\epsilon_{0}(\vk)\, {\bold I}_{\rm 2x2}+  \vd(\vk) . \boldsymbol{\sigma} ,
\label{Hamilton2by2}
\end{equation}
where $\boldsymbol{\sigma}=(\sigma_x,\sigma_y,\sigma_z)$ is the vector of standard Pauli matrices representing some internal isospin degree of freedom. This degree of freedom can be either a real spin, the sublattice index (A and B sublattices of graphene), an orbital index ($s$ and $p$ orbitals defined on the same site). The details of the coupling is described by the vector $\vd(\vk)=(d_x(\vk),d_y(\vk),d_z(\vk))$ of periodic functions of $\vk$. The structure of the Bloch Hamiltonian $\mathcal{H}(\vk)$ is constrained by the symmetries of the problem. 

All the information about the topology of wave functions is encoded in four real and periodic functions of the momentum, $(\epsilon_0(\vk),d_x(\vk),d_y(\vk),d_z(\vk))$, all defined on the whole Brillouin zone $T^2$. The function $\epsilon_0(\vk)$ simply shifts the eigenvalues without affecting the eigenstates, and therefore it has no effect on the topological properties of the material. Nevertheless this function is very important $\epsilon_0(\vk)$ because it enters the spectrum dispersion and determination of the position of the Fermi level.

\bigskip

{\it Example:} The Haldane model \cite{Haldane:1988} is defined by the functions:
\begin{align}
\epsilon_0(\vk) & =2 t_2 \cos (\phi)   \sum_{i=1}^{3}       \cos(\vk . \vb_i) ,\\
d_x (\vk)  & =  t \, \Re \gamma(\vk) =  t \sum_{\alpha=1}^{3} \cos(\vk . \boldsymbol{\delta}_\alpha) ,\\
d_y (\vk)  & =  t \, \Im \gamma(\vk) =  t \sum_{\alpha=1}^{3}  \sin(\vk . \boldsymbol{\delta}_\alpha)  ,\\ 
 d_z (\vk)  & =  M_1 + 2 t_2 \sin (\phi)   \sum_{i=1}^{3}    \sin(\vk . \vb_i)  ,
\end{align}
where the 2 dimensional momentum $\vk$ lives in the first Brillouin zone $T^2$. As seen in chapter \ref{chapter0}, this model breaks simultaneously two symmetries, inversion and time-reversal. The terms $\epsilon_{0} {\bold I}_{\rm 2x2},\sigma_x d_x(\vk),\sigma_y d_y(\vk)$ (which describes "unperturbated" graphene) are invariant under both $\mathcal{T}$ and $\mathcal{P}$. In contrast the $d_z(\vk)\sigma_z$ term is not invariant under $\mathcal{T}$ due to the flux term $t_2 \sin(\phi)$ term {\cite{Haldane:1988}}, and not invariant under $\mathcal{P}$ due to the on-site staggered term $M_1$ \cite{Semenoff:1984}.

\subsection{Spectrum and wave functions}

The general Hamiltonian Eq.(\ref{Hamilton2by2}) can be easily diagonalized. The band structure consists of an upper ($\alpha=+$) and a lower ($\alpha=-$) bands:
\begin{equation}
E_{\alpha=\pm}(\vk)=\epsilon_0(\vk) \pm |\vd(\vk)|, 
\end{equation}
the corresponding wave functions being the spinors:
\begin{equation}
\Phi_+(\vk) =
\begin{pmatrix}
u_{+1}(\vk) \\ 
u_{+2}(\vk)
\end{pmatrix}=\begin{pmatrix}
\cos \frac{\theta}{2} e^{-i\phi/2}\\ 
\sin \frac{\theta}{2} e^{i\phi/2}
\end{pmatrix},  
\end{equation}
in the upper band, and 
\begin{equation}
\Phi_- (\vk)=
\begin{pmatrix}
u_{-1}(\vk) \\ 
u_{-2}(\vk)
\end{pmatrix}=\begin{pmatrix}
\sin \frac{\theta}{2} e^{-i\phi/2}\\ 
-\cos \frac{\theta}{2} e^{i\phi/2}
\end{pmatrix},  
\end{equation}
in the lower band. The $\vk$-dependent quantities $\theta=\theta_\vk$ and $\phi=\phi_\vk$ are the spherical coordinate angles of the unit vector:
\begin{equation}
\dhat (\vk) = \frac{\vd(\vk)}{|\vd(\vk)|}=
\begin{pmatrix}
\cos \phi \sin \theta \\ 
\sin \phi \sin \theta \\
\cos \theta
\end{pmatrix},
\label{dspherique}
\end{equation}
which resides on the unit sphere $S^2$ while $\vk$ spans the $d$-dimensional toroidal Brillouin zone $T^d$. The mapping $\vk \rightarrow \dhat (\vk) = \vd(\vk)/|\vd(\vk)|$ is essential and captures the topological properties of the Hamiltonian Eq.(\ref{Hamilton2by2}). We assume that the system is insulating so $|\vd(\vk)| \neq 0$ everywhere and the mapping is well defined in the whole FBZ.

\subsection{Berry phase and anomalous velocity} 
We can interpret $\vk$ as a parameter that we can vary along a loop drawn in the FBZ and limit ourselves to $d=2$. Along such a loop, the spinor will acquire a Berry phase which is the circulation of the Berry vector potential, also called Berry connection defined by 
\begin{equation}
\vA_\alpha (\vk)= i \sum_{a=1}^{2} u_{\alpha a} \nabla_{\vk} u_{\alpha a},
\label{berry}
\end{equation}
in each band $\alpha=\pm 1$. The Berry curvature is the curl of the Berry connection:
\begin{equation}
\vB_\alpha(\vk)=\nabla_\vk \wedge \vA_\alpha(\vk).
\label{berrycurvature}
\end{equation}
The flux of $\vB_\alpha(\vk)$ through the whole FBZ (torus $T^2$), 
\begin{equation}
C_\alpha =\frac{1}{2 \pi} \int d\vk \, \vB_\alpha(\vk),
\label{FluxBerry}
\end{equation}
is called the Chern number of the band $\alpha$. The Chern number is defined from the projectors on the lower (occupied) band. For the Haldane model, the valence and the conduction bands have 
a non zero Chern number ($\pm 1$).

\bigskip

Finally, from the Bloch Hamiltonian's expression (\ref{Hamilton2by2}) one can define the components $j_{i=x,y}$ of the current operator:
\begin{equation}
j_{i}=\frac{\partial \mathcal{H}(\vk)}{\partial k_i}= \frac{\partial\epsilon_{0}(\vk) }{\partial k_i} {\bold I}_{\rm 2x2}+  \sum_{j}^{3} \frac{\partial \vd(\vk)}{\partial k_i} .  \boldsymbol{\sigma} ,
\label{AnomalousCurrent}
\end{equation}
which are $2\times2$ matrices for a two-band insulator. The second term, and in particular its non-diagonal elements, are responsible for Berry phase effects and eventually finite Hall response.

\section{Topological invariant \label{sectionChern}}
The electromagnetic response of 2D two-band insulators can be computed using the Kubo formalism and the expressions of the anomalous currents Eq. (\ref{AnomalousCurrent}). In $d=2$ space dimensions, the Hall conductance is exactly given by the Chern number of the lower band. This is also the winding number of the mapping $\vk \rightarrow \dhat (\vk) = \vd(\vk)/|\vd(\vk)|$, which explains geometrically the quantization of the Hall conductance in such two-band model. This is reminiscent of the Thouless-Kohmoto-Nightingale-den Nijs (TKKN) invariant for quantum Hall systems \cite{Thouless:1982}.

\subsection{Hall conductivity as a winding number}

The Hall conductivity can be calculated from Kubo formalism as \cite{Qi:2006}:
\begin{equation}
\sigma_{xy} =\frac{e^2}{4 \pi h}  \int d^2 \vk  (f_+(\vk) -  f_-(\vk))   \left(  \frac{ \partial \dhat(\vk)}{\partial k_x}     \times  \frac{ \partial \dhat(\vk)}{\partial k_y}   \right) .   \dhat(\vk) ,
\label{Hallconductance}
\end{equation}
where $f_\pm (\vk)$ are the occupation numbers of the conduction and valence bands. It is assumed that the Fermi level lies in the bulk gap. Hence at zero temperature, where $f_- = 1$ and $f_+ =0$, we have the relation:
\begin{equation}
\sigma_{xy} =\frac{e^2}{h}  n_w ,
\label{Winding1}
\end{equation}
where $n_w$ is the winding number (or Pontryagin index) of the mapping $\vk \rightarrow \dhat (\vk) = \vd(\vk)/|\vd(\vk)|$ between the Brillouin zone (torus $T^2$) and the unit sphere ($S^1$):
\begin{equation}
n_w =\frac{1}{4 \pi}  \int d^2 \vk   \left(  \frac{ \partial \dhat (\vk)}{\partial k_x}     \times  \frac{ \partial \dhat (\vk)}{\partial k_y}   \right) .   \dhat .
\label{Winding2}
\end{equation}
In contrast to the Berry phase, this number is directly constructed from the parameters $\vd(\vk)$ of the Hamiltonian Eq. \ref{Hamilton2by2} (rather than from its eigenstates). This winding number is an integer that counts the number of times the unit vector $\dhat(\vk) $ wraps around the whole sphere $S^2$ while $\vk$ is spanning the whole Brillouin zone $T^2$. In accordance with general classifications, there is a single number that characterizes the general structure of wave functions globally in $\vk$-space. This number is a relative integer, and it measures the charge Hall conductance in units of $e^2/h$. To change $n_w$ it is necessary to change the parameter of the bulk Hamiltonian $\dhat(\vk)$ in such a way that the bulk gap closes.

{\it Interpretation as a magnetic textures}
We can see the field $\dhat(\vk)$ as some texture in momentum space. A terminology has been establish in real space to distinguish topological defects like skyrmions and merons. Those fields carry a topological charge which is just the winding number defined above provided momentum $\vk$ is replaced by position and also the FBZ is replaced by the manifold upon which the texture resides.

\subsection{Topological phase transition and band inversion}

{\it Haldane model.} Let us calculate this winding number for the simple model of massive Dirac fermions introduced previously. We use the parametrization Eq. (\ref{dspherique}) to rewrite the winding number as: 
\begin{align}
n_w &=\frac{1}{4 \pi}  \int d^2 \vk   \sin \theta \left(  \frac{\partial \theta}{\partial k_x} \frac{\partial \phi}{\partial k_y} -  \frac{\partial \phi}{\partial k_x} \frac{\partial \theta}{\partial k_y}  \right) .   \dhat  ,\\
&= - \frac{1}{4 \pi}  \int d^2 \vk   \nabla \wedge \left(  \cos \theta \nabla \phi   \right) ,
\label{Winding3}
\end{align}
which in principle should be always zero because we integrate over a torus. Finite values can arise from singularities in $ \nabla \phi  $ that are always located at the poles $\theta=0$ and $\theta=\pi$. Hence:
\begin{equation}
n_w =- \frac{1}{4 \pi}  \int \cos \theta \nabla \phi   . d\vl   ,
\label{Winding4}
\end{equation}
where the integral is taken along loops encircling the poles. 

\medskip

For graphene, the poles are reached when $\vk$ is at the Dirac points. Then the sign of $d_z(\vk=\pm \vK)$ indicates whether the north or south pole has been reached. For the Semenoff mass, we have $d_z(\vk=\pm \vK) =M_1) $ in both valleys. Then we have to notice that $\vd = (\xi k_x , k_y,M_1) $ accumulates opposite phases while winding around the same pole (due to the presence of the valley index $\xi$). For the Haldane mass, one has $d_z(\vk=\pm \vK) =M_H \xi)$ which means that the accumulated phases (at south and north poles respectively) add up and finally:
\begin{equation}
n_w =- \frac{1}{4 \pi} \left( 2\pi + 2\pi\right) sign(M_H) =-sign(M_H).
\end{equation} 
In fact for each valley, the winding of the angle $\phi$ is $2 \pi \xi$ and $\cos \theta = Sign (M_\xi)$ where $M_\xi$ is the mass in valley $\xi$. Therefore one has
\begin{equation}
n_w =- \frac{1}{4 \pi} \sum_{\xi=\pm 1} 2 \pi \xi sign (M_\xi)=\frac{1}{2} \left(  sign(M_-) - sign(M_+) \right),
\end{equation} 
which explains why the global winding number is zero when the masses are equal in both valleys, and why it is $n_w =\pm 1$ in the Haldane phase characterized by a band inversion. This has been formulated in a more elegant and general way in Ref. \cite{Sticlet:2012}.

\medskip

Let us now consider the example of graphene in presence of some inversion breaking and time-reversal breaking terms. So the mass matrix is $(M_1 - 3 \sqrt{3} t_2 \sin (\phi) \xi ) \sigma_z$ implying that the gap can close for $M_1 =\xi 3 \sqrt{3} t_2 \sin \phi$ in one valley (labelled by $\xi=\pm 1$). This equality signals a one-electron topological quantum transition separating a QH insulator for a trivial atomic insulator. Finally we would like to make a comment on the terminology. This type of phase transition is purely a change between two one-electron Hamiltonian. It has in particular nothing to do with topological order defined by Wen. In particular the transition discussed here is not a transition between two topological orders. It is rather a transition between two band-insulators having distinct topological invariants (which characterize the winding of one-electron wave functions).

\section{Chiral edge states \label{sectionEdge}}

Here we describe a rather general phenomenon occurring when an insulating (gapped) medium develops a region where the gap changes sign. Fermionic zero energy modes can show up in the gapless interfacial region, and get confined between two fully gapped bulk regions. Two different situations are to be contrasted. First one might consider a mass kink without change of the topological invariant, for instance between two Semenov insulators, or two Haldane insulators (in the terminology introduced in chapter \ref{chapter1}) having opposite masses. Then the edge states exist but they are generally not protected. In contrast, the edge states residing at the interface between a Semenov phase and an Haldane phase (two topologically distinct insulators) is protected. We specialize our examples to 2D insulators so that the edge modes are running along 1D interfaces. Nevertheless the idea is rather general and valid for $D-1$ dimensional surface gapless modes emerging at interfaces between $D$ dimensional gapped phases. This physics is reminiscent of the Jackiw-Rebbi model introduced in field theory \cite{Jackiw:1976} and of the physics of solitons in polyacetylene (D=1) \cite{SSH:1979,SSH:1980}.

\subsection{Interface between topologically distinct insulators \label{HaldaneSemenov}}

\begin{figure}
\begin{center}
\includegraphics[width=6cm]{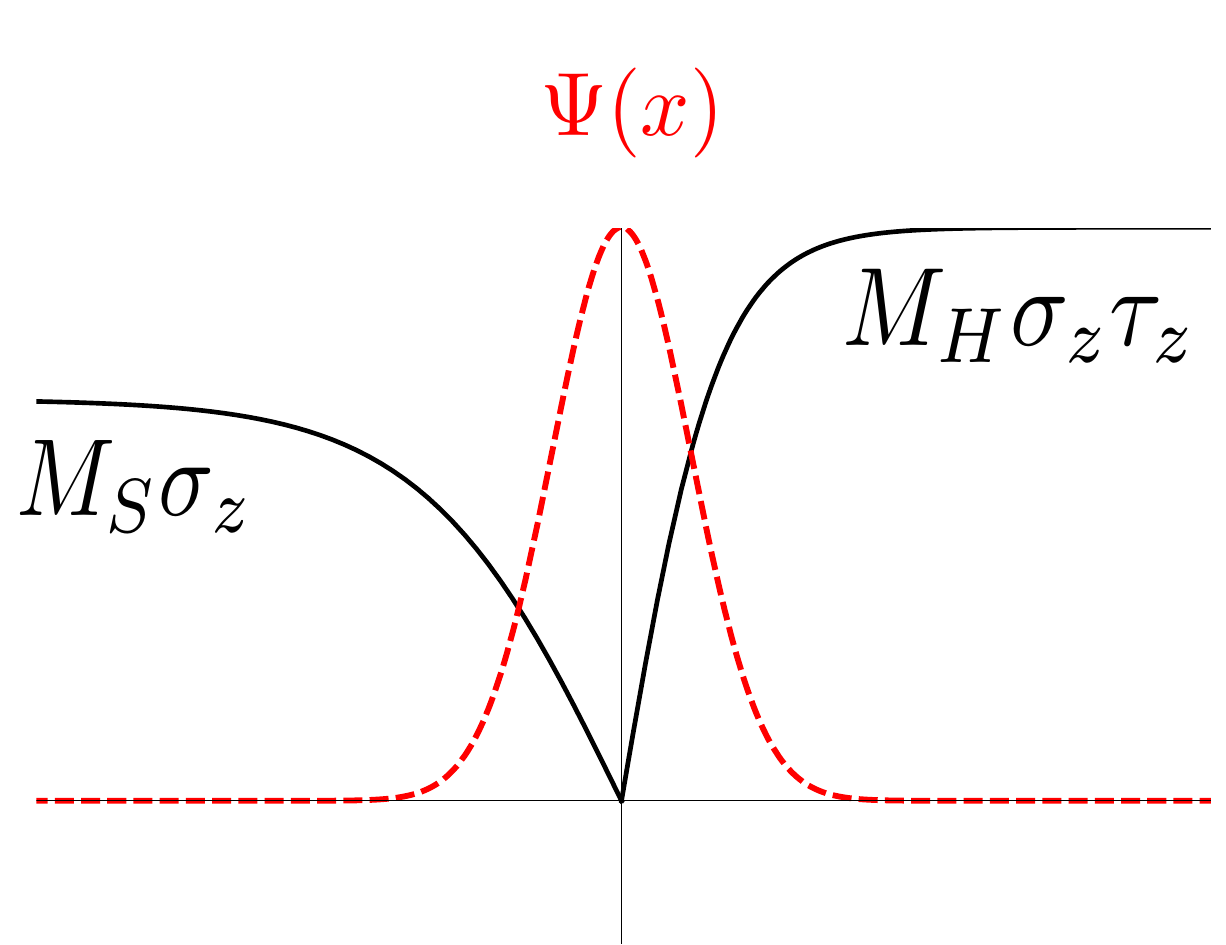}
\includegraphics[width=6cm]{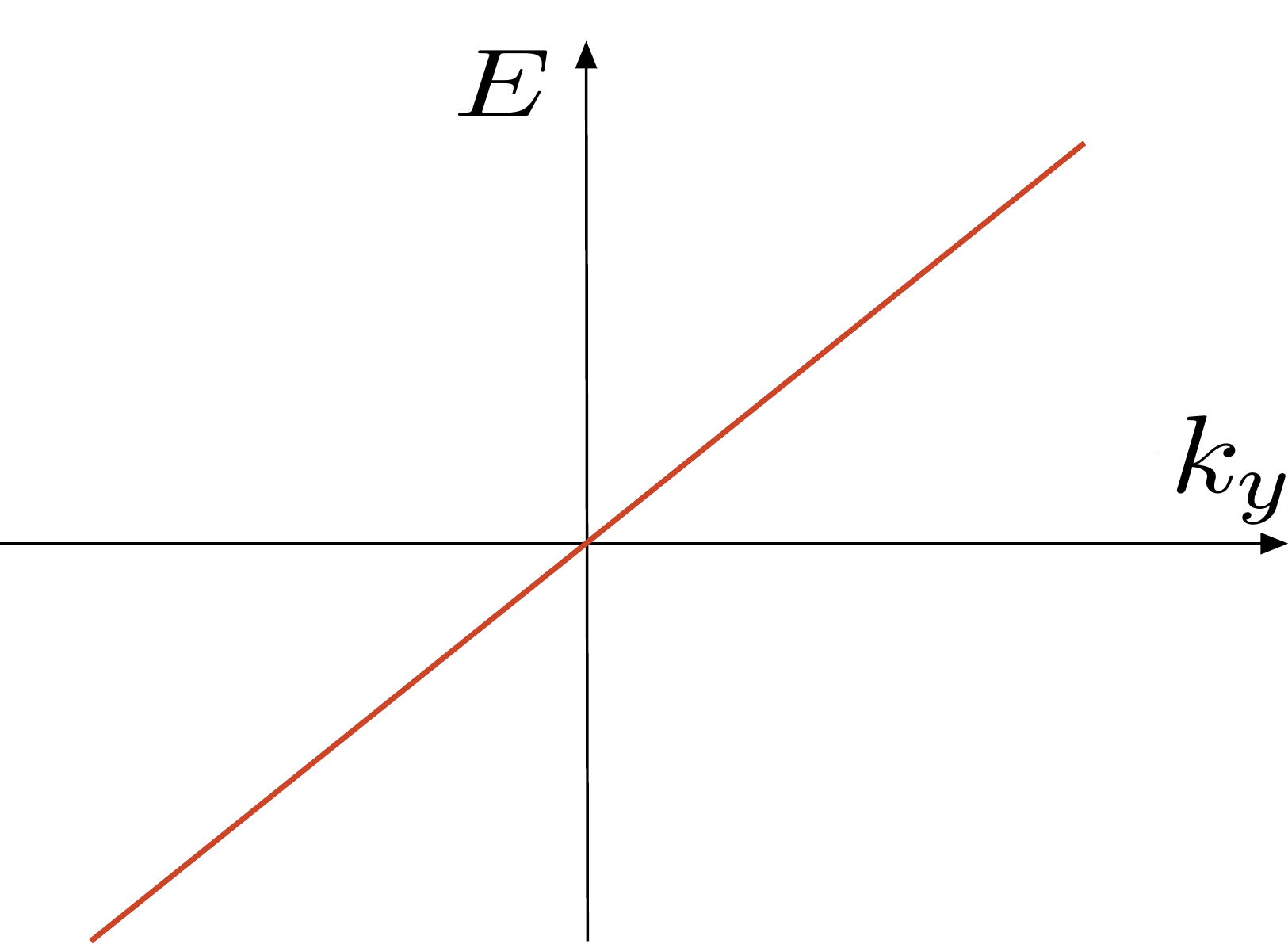}
\caption{{\it Left panel:} Interface between the Haldane insulator ($x>0$) and the Semenov insulator ($x<0$). There is always a zero energy bound state located near $x=0$ (red dashed line). {\it Right panel:} dispersion $E(k_y)={\rm sign}(M_H) \hbar v_F k_y $ of the chiral edge mode.}
\end{center}
\label{FigHaldaneSemenovWall}
\end{figure}

We assume that the half-plane $x<0$ is filled with a "Semenov" (inversion-breaking) insulator while an "Haldane" (time-reversal breaking) insulator occupies the half-plane $x>0$. In principle one should define this heterojunction on the lattice by varying the parameters of the Haldane model (namely the on-site mass and the chiral phase $\phi$) near the interface. Since we are mainly interested in eventual zero modes confined near the interface $x=0$, we use the low energy effective model valid near the Dirac points (energies smaller than the bandwidth $t$).  Using the translational invariance along the $y-$direction, the wave equation reads: 
\begin{equation}
\left( -i \hbar v_F \sigma_x \tau_z \partial_x +\hbar v_F k_y  \sigma_y + M(x)  \right) \Psi = E \Psi,
\label{waveeq}
\end{equation}
with $M(x)=M_S \Theta(-x) \sigma_z + M_H \Theta(x) \sigma_z \tau_z$. In fact we can even consider a more general shape by replacing the Heaviside functions $\Theta(x)$ by smooth functions interpolating between zero for negative arguments and unity for large positive arguments. Nevertheless the sharp step model is accurate provided the length scale for the variation of the lattice parameters is smaller than the extension of the eventual edge state, namely $\hbar v_F /{\rm max}(M_{S},M_{H})$.   

Let us first show there is always a solution at $E=0$ and $k_y=0$ by solving the equation:
\begin{equation}
\hbar v_F \partial_x \Psi = -i \sigma_x \tau_z  M(x) \Psi ,
\end{equation}
obtained by multiplying both sides of Eq. (\ref{waveeq}) by $i \sigma_x \tau_z$.
For $x>0$, we have:
\begin{equation}
\hbar v_F \partial_x \Psi = - \sigma_y   M_H \Psi ,
\end{equation}
and the bounded solution (decaying at $x \rightarrow \infty$) is the eigenstate of $\sigma_y$ with eigenvalue sign$(M_{H})$

For $x<0$, there is an additional valley matrix $\tau_z$ in the wave equation:
\begin{equation}
\hbar v_F \partial_x \Psi = - \sigma_y \tau_z   M_S \Psi ,
\end{equation}
and the corresponding bounded solution is the eigenstate of $\sigma_y$ with eigenvalue -sign$(\xi M_S)$. So the matching is possible, and there is a zero mode at the boundary, only if the two solutions above correspond to the same eigenvalue of $\sigma_y$, namely if
\begin{equation}
sign(M_{H})=-sign(\xi M_S).
\end{equation}
For any choice of the masses, this equality is always valid in one valley which is fixed by the relative signs of $M_{H}$ and $M_{S}$. Therefore one obtains a zero mode which {\it is polarized in the valley $\xi=-sign(M_{S}M_{H})$}. 

Now in order to obtain the wave function and dispersion $E(k_{y})$ of this edge mode, let us restore finite energy $E$ and parallel momentum $k_y$ in Eq.(\ref{waveeq}). Without further calculation one notice that the zero mode at $k_y=0$ is also eigenstate of $\hbar v_F k_y  \sigma_y$, and therefore its expression is still valid at finite energy and momentum with the dispersion:
\begin{equation}
E=-sign(\xi M_{S}) \hbar v_F k_y =sign(M_{H}) \hbar v_F k_y .
\end{equation}
The edge mode is chiral and shows up in the valley that is experiencing a mass inversion at the interface. In the limit of large $M_S$, the Semenov insulator can represent the vacuum. By reproducing this calculation for various orientation of the interface it is easy to demonstrate that the Haldane insulator is surrounded by a 1D edge chiral edge mode that circulates clockwise if $sign(M_S M_H)$ is positive, and anti-clowise for negative $M_S M_H$. Note that if we assume that the vacuum is represented by a large positive $M_S$, then the sign of $M_S M_H$ is simply the sign of $M_H=-3\sqrt{3} t_2 \sin(\phi)$ which is set by the chirality of the flux pattern in the microscopic Haldane model (see chapter \ref{chapter1}).

\subsection{Kink in the Haldane mass}
As was the previous one, this section might be a bit academic since it requires to make a junction between two Haldane insulators with opposite chiralities while there is not yet any experimental evidence of a Haldane phase in graphene. Nevertheless we think one can learn a lot from those simple toy models. 

\begin{figure}
\begin{center}
\includegraphics[width=6cm]{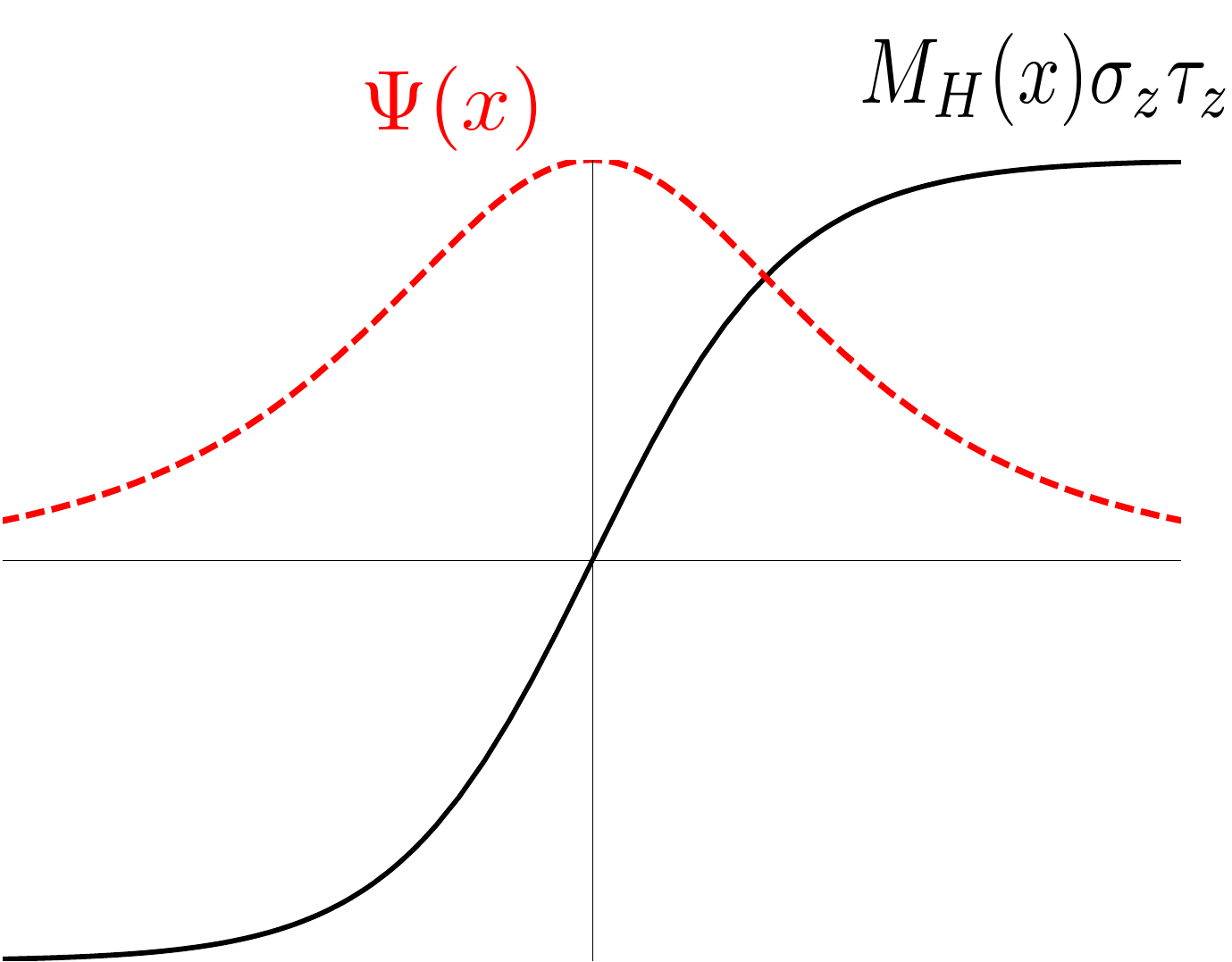}
\includegraphics[width=6cm]{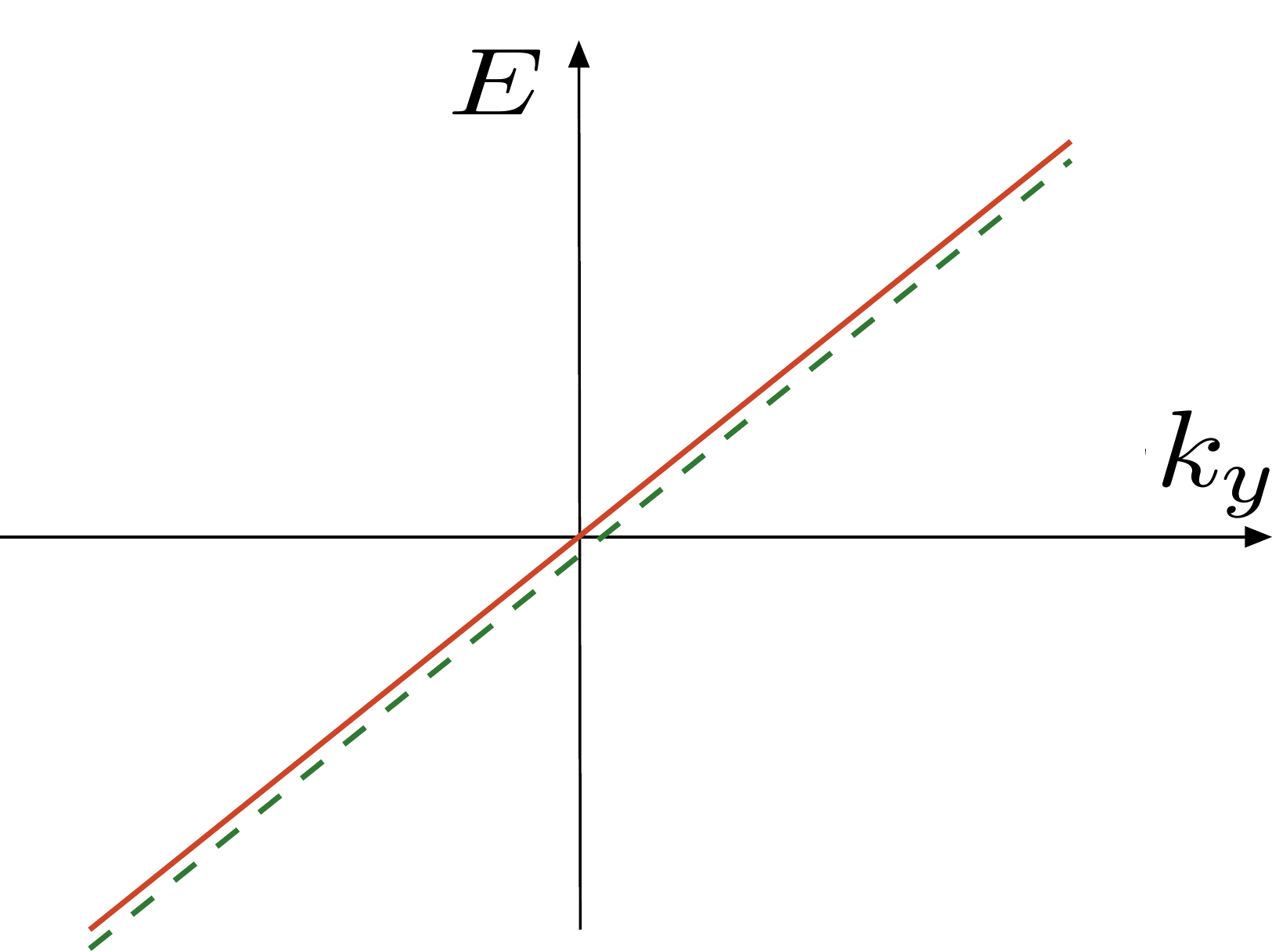}
\caption{{\it Left panel:} Kink in the Haldane mass with the sign change occurring at $x=0$. There is always twofold degenerated zero energy bound states located near $x=0$ (red dashed line). {\it Right panel:} dispersion $E(k_y)={\rm sign}(M_H) \hbar v_F k_y $ of the two independent chiral edge modes (solid and dashed curves respectively).}
\end{center}
\label{FigKinkHaldane}
\end{figure}
The full wave equation for the Haldane kink reads: 
\begin{equation}
\left( -i \hbar v_F \sigma_x \tau_z \partial_x +\hbar v_F k_y  \sigma_y + M_H(x) \sigma_z \tau_z  \right) \Psi = E \Psi,
\label{waveeqKinkHaldane}
\end{equation}
where $M_H(x)$ is a real monotonic function describing a kink with $M_H(\infty)$ positive and $M_H(-\infty)$ negative hereafter (the opposite case can be treated similarly). We take the origin $x=0$ where $M(x)$ has its zero. We expect that a bound state might show up near $x=0$ because the insulator becomes "locally" gapless there. 

We first look for a $E=0$ solution at $k_y=0$ by solving the equation: 
\begin{equation}
 i \hbar v_F \sigma_x \tau_z \partial_x \Psi  = M_H(x) \sigma_z \tau_z  \Psi .
\label{waveeqKinkHaldane2}
\end{equation} 
By multiplying each side by $-i \sigma_x \tau_z$, it is obtained:
\begin{equation}
\hbar v_F  \partial_x \Psi  = - M_H(x) \sigma_y \Psi ,
\label{waveeqKinkHaldane3}
\end{equation} 
which has the solution:
\begin{align}
\Psi(x) & =\exp \left( - \int_{0}^{x} dx' M_H(x')/\hbar v_F \right) | \sigma_y=+1 \rangle \\
 &= \exp \left( - \int_{0}^{x} dx' M_H(x')/\hbar v_F \right) 
 \left[
a
\begin{pmatrix}
1 \\ 
i\\
0\\
0\\
\end{pmatrix}
+
b
\begin{pmatrix}
0 \\ 
0\\
1\\
i\\
\end{pmatrix}
\right]
\label{waveeqKinkHaldaneSolution}
\end{align} 
Hence there is a twofold degenerate zero mode at $k_y=0$.

Now we can restore a finite transverse momentum $k_y$ and observe that the above solution is an eigenmode of $\hbar v_F \sigma_y$ with energy $E=\hbar v_F k_y$. The two degenerate chiral zero modes Eq.(\ref{waveeqKinkHaldaneSolution}) yield two degenerate chiral modes propagating in the same direction along $y$-axis. This is consistent with the fact that the Haldane model breaks time-reversal symmetry. We can understand the Haldane kink as two remote Haldane insulators (with opposite chiralities) that would have been brought in contact together adiabatically. After such a process one would have to modes running in the same direction along the interface considered.

\subsection{Kink in the Semenov mass}
\begin{figure}
{\includegraphics[width=6cm]{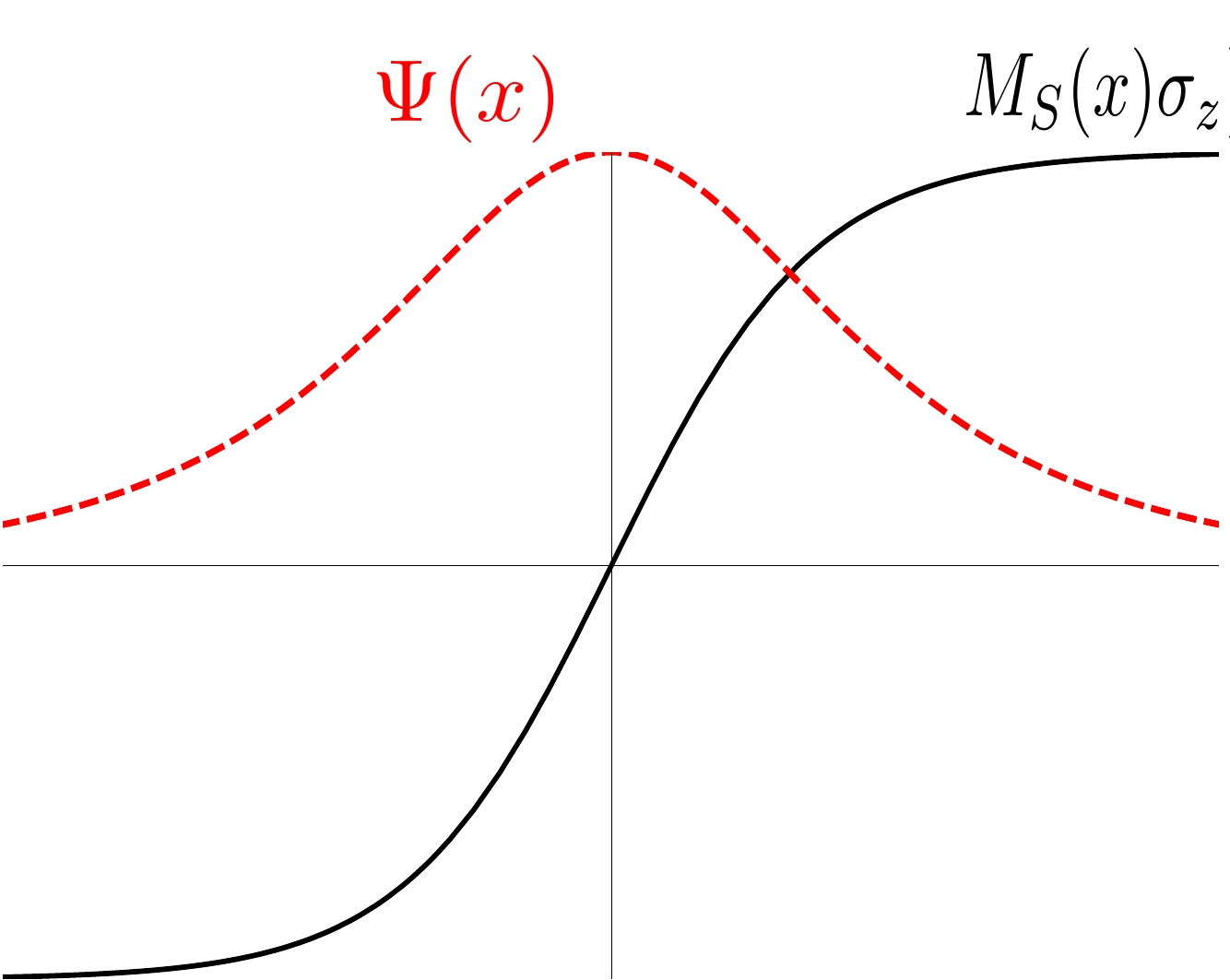}}
{\includegraphics[width=6cm]{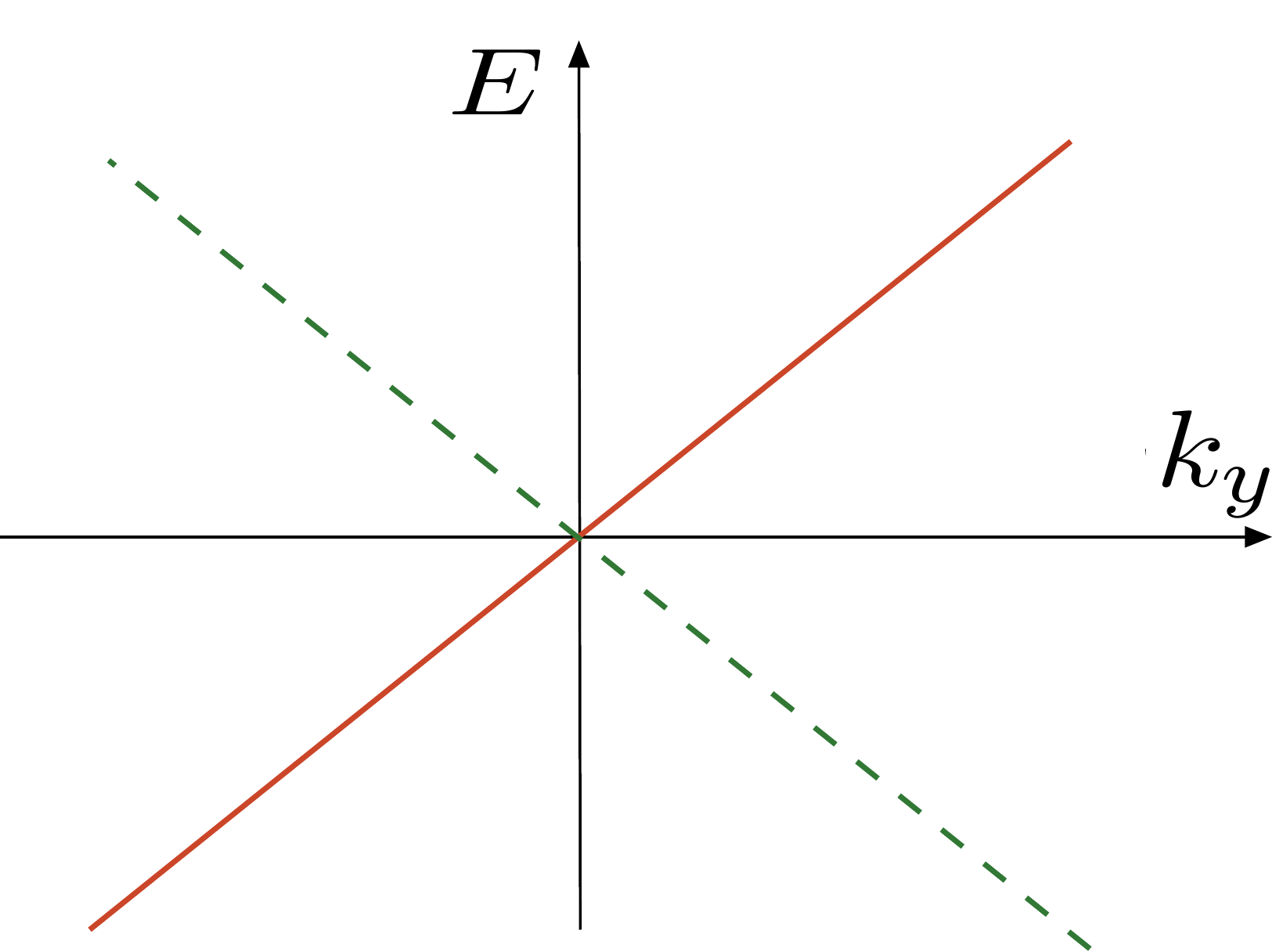}}
\caption{{\it Left panel:} Kink in the Semenov mass $M(x) \sigma_z$ with the sign change occurring at $x=0$. There is always twofold degenerated zero energy bound states located near $x=0$ (red dashed line). {\it Right panel:} dispersion $E(k_y)=\pm \hbar v_F k_y $ of the two counter-propagating edge modes (solid and dashed curves respectively). This counter propagation is the natural consequence of the time-reversal invariance of the system.}

\label{FigKinkSemenov}
\end{figure}

One can easily reproduce the similar analysis for a kink of the Semenoff mass by solving the wave equation: 
\begin{equation}
\left( -i \hbar v_F \sigma_x \tau_z \partial_x +\hbar v_F k_y  \sigma_y + M_S(x) \sigma_z  \right) \Psi = E \Psi,
\label{waveeqKinkSemenov1}
\end{equation}
where $M_S(x)$ is a real function satisfying $M_S(0)=0$, $M_S(\infty)>0$ and $M_S(-\infty)<0$. The equation for the eventual zero energy mode at $k_y=0$ is then:
\begin{equation}
\hbar v_F  \partial_x \Psi  = -M_S(x)  \sigma_y \tau_z \Psi ,
\label{waveeqKinkSemenov2}
\end{equation} 
whose solution reads:
\begin{align}
\Psi(x) & =\exp \left(- \int_{0}^{x} dx' M_S(x')/\hbar v_F \right) | \sigma_y \tau_z=+1 \rangle \\
 &= \exp \left( - \int_{0}^{x} dx' M_S(x')/\hbar v_F \right) 
 \left(
a
\begin{pmatrix}
1 \\ 
i\\
0\\
0\\
\end{pmatrix}
+
b
\begin{pmatrix}
0 \\ 
0\\
1\\
-i\\
\end{pmatrix}
\right).
\label{waveeqKinkSemenovSolution}
\end{align}
As  a major difference with the Haldane kink, the two parts of the wave function leads to opposite chiralities when a finite $k_y$ is restored. This is because they correspond to eigenmodes of $\sigma_y$ with opposite eigenvalues $\pm 1$. This is consistent with the global time-reversal symmetry of the system.

\section*{Conclusion}
We have taken advantage of the conceptual simplicity of CIs to introduce the general ideas topological invariants and topologically protected gapless edge modes using the Haldane model as a guiding example. At the interface between two insulators with opposite masses (same matrix but sign inversion through the interface), or between two topologically distinct insulators, zero modes generically show up and form an interfacial metal. When the two insulators have distinct topological invariants, the metallic (gapless) of this interfacial electronic system (holographic) is topologically protected . This is quite similar to the situation of the chiral edge states of the integer QHE protected by the finite TKKN invariant. The ideas discussed in this chapter will be used in the following to understand other topological states, like the Quantum Spin Hall (QSH) state (see Chapter \ref{chapter2}) and Floquet topological insulators (see chapter \ref{chapter5}). 

\chapter{Topological insulators \label{chapter2}}

In the absence of spin-orbit coupling, the Dirac points of graphene are protected by the combination of two fundamental discrete symmetries: time-reversal and spacial inversion (chapter \ref{chapter0}). In 2005, C.L. Kane and E.G. Mele demonstrated that the situation is drastically changed when the spin is coupled to electronic motion: intrinsic spin-orbit coupling does open a gap at the Dirac points without breaking any of those fundamental symmetries \cite{Kane:2005a,Kane:2005b}. The resulting insulator, the so-called Quantum Spin Hall (QSH), is a novel state of electronic matter that cannot be adiabatically connected to a trivial atomic insulator without closing (and re-opening) the bulk gap. The QSH state is distinguished from ordinary band insulators by the presence of a one-dimensional metal along its edge \cite{Kane:2005a,Kane:2005b} which is topologically protected by a Z$_2$ topological invariant. The nonchiral QSH edge states are also different from the chiral edge states of the Quantum Hall insulators or Chern insulators (described in the previous chapter), thereby providing a new class of one-dimensional (1D) conductors. Interestingly the direction of the spin of the 1D charge carriers is tied to their direction of motion. Such conductors are protected from single-particle backscattering (and Anderson localization) by time-reversal symmetry $\mathcal T$. 

\medskip

Unfortunately the QSH state is extremely difficult to observe in graphene due to the actual weakness of the spin-orbit interaction  \cite{HuertasPRB:2006,MinPRB:2006}. In 2006, Bernevig, Hughes and Zhang (BHZ) predicted that CdTe/HgTe/CdTe quantum wells should host such a QSH state in their inverted regime \cite{Bernevig:2006}. Their prediction was soon followed by the experimental observation of conducting edge states by the group led by Laurens Molenkamp  \cite{Konig:2007,Roth:2009}. This experimental confirmation has triggered a great deal of excitation in the condensed matter community \cite{KaneRMP:2011,QiRMP:2011,KonigJPSJ:2008,QiPhysToday:2010}. 

\medskip

This chapter is organized as follows. We start by introducing the Kane-Mele model for graphene and the BHZ model for HgTe/CdTe heterostructures at the level of idealistic spin-conserving Hamiltonians (sections \ref{sectionKM} and \ref{sectionBHZ}): then the QSH state consists in two time-reversed Chern insulators and therefore exhibits spin filtered counterpropagating edge modes. At this level of approximation, backscattering is forbidden owing to spin-conservation. Finally and most importantly, we emphasize that the QSH edge states are actually robust {\it even in presence of spin-mixing terms} owing to the protection by a bulk $Z_2$ topological invariant (section \ref{sectionedge}). This invariant originates from the Kramers degeneracy property applied to time-reversal invariant band structures of fermions.   

\section{Kane-Mele model of graphene \label{sectionKM}}
The Haldane model for spinless fermions on the honeycomb lattice (chapter \ref{chapter0}) and other Chern insulator models (chapter \ref{chapter1}) break time-reversal symmetry and have interesting topological properties akin to the Integer Quantum Hall Effect \cite{Haldane:1988}. Nevertheless such Chern insulators, remain difficult to realize experimentally because some highly nontrivial internal magnetic background is required. In 2005, C.L. Kane and E.G. Mele proposed a generalization of the Haldane model that respects time-reversal invariance and includes the spin via the spin-orbit interaction. Their idea launched the field of time-reversal invariant topological insulators which has known a rapid expansion since then \cite{KaneRMP:2011,QiRMP:2011,KonigJPSJ:2008,QiPhysToday:2010}.

\subsection{Intrinsic spin-orbit coupling}

We first discuss the idealized situation of a spin-orbit coupling that still conserves one component of the electronic spin. In their seminal paper, C.L. Kane and E.G. Mele introduced the following lattice model for spinfull electrons on the honeycomb lattice  \cite{Kane:2005a}:  
\begin{equation}
\label{KaneMeleHamiltonian}
H=  t \sum_{\langle i,j \rangle}    c^\dagger_{i\alpha} c_{j\alpha} + i  t_2 \sum_{\langle \langle i,j \rangle \rangle}  \nu_{ij}   c^\dagger_{i\alpha} (s_{z})_{\alpha \beta} c_{j\beta},
\end{equation}
where the Pauli matrix $s_z$ refers to the physical spin of electrons, and the summation over repeated spin index ($\alpha,\beta$) is implied. The next-nearest neighbor (NNN) hopping term $i t_2 \nu_{ij} s_z$ describes a spin-orbit coupling between the spin direction $s_z=\pm 1$ (units of $\hbar/2$) and the chirality $\nu_{ij}$ of the circulating electrons. This can be seen as a $L.S$ coupling where the "orbital momentum" $L$ would be associated with the chirality. One can develop a very simple local picture for this spin-orbit coupling. 


\medskip

{\it Two copies of the Haldane insulator.} Since $[H,s_z]=0$, the model Eq.(\ref{KaneMeleHamiltonian}) can be decoupled into two subsystems for spin up ($s_z =1$) and spin-down ($s_z=-1$) respectively. The Hamiltonian for spin-up (resp. spin-down) electrons is the Haldane Hamiltonian Eq.(\ref{HaldaneHamiltonian}) with $\phi=\pi/2$ (resp. $\phi=-\pi/2$). Hence many properties can be deduced from our knowledge of the Haldane model for spinless fermions (chapter \ref{chapter1}).

\medskip

\begin{figure}
\begin{center}
\includegraphics[width=12cm]{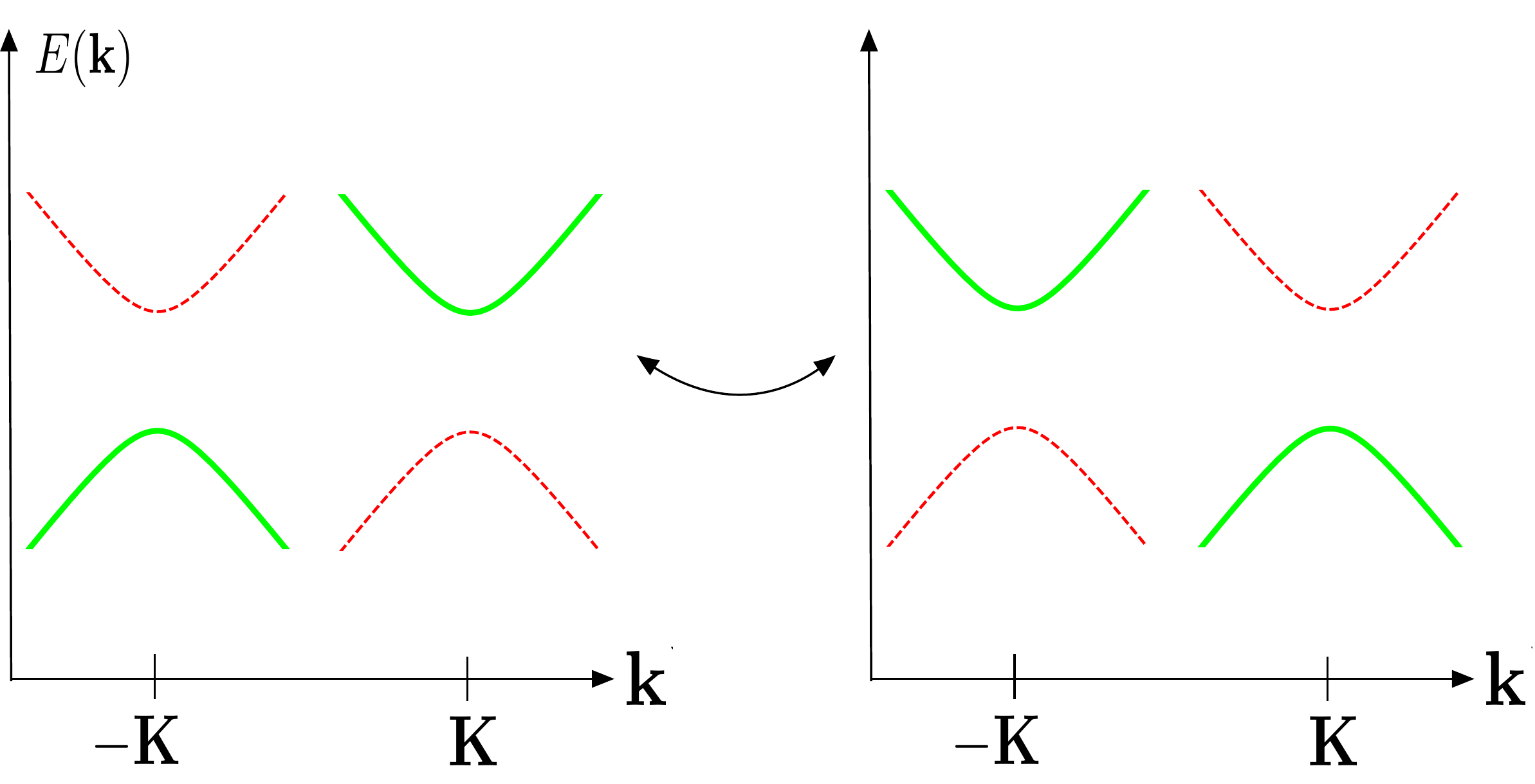}
\caption{Low energy dispersion for the spinfull Kane-Mele insulator (consisting in two time-reversed copies of a Haldane insulator).}
\end{center}
\label{FigKaneMeleBands}
\end{figure}

{\it Insulator in the bulk.} Firstly, the system is gapped in the bulk. Indeed the low-energy theory of the lattice Hamiltonian Eq.(\ref{KaneMeleHamiltonian}) is directly derived from Eq.(\ref{HaldaneTerm}):
 \begin{equation}
\mathcal{H}_{\rm so} = \Delta_{\rm so} \sigma_z \tau_z  s_z,
\label{HamiltonianKMEffective}
\end{equation}
where $\Delta_{\rm so}=-3\sqrt{3} t_2$. This perturbation anticommutes with the kinetic Hamiltonian $\mathcal{H}_0$, and therefore opens a gap at the Dirac points. For each spin specy, there is a mass inversion between the two valleys $m_{\vK \alpha}=-m_{-\vK \alpha}$ as in the Haldane model. Nevertheless the global electronic system is a time-reversal invariant insulator in the bulk because $m_{\vK \alpha}=m_{-\vK -\alpha}$ (\ref{FigKaneMeleBands}).

More formally one can consider the action of the time-reversal operator $\mathcal T$ on the perturbation Eq.(\ref{HamiltonianKMEffective}). For spinless electrons on the honeycomb lattice, this operator is $\mathcal T=\tau_x K$ where $\tau_x$ switches the valleys and $K$ is the complex conjugation (see chapter \ref{chapter0}). For spinfull electrons it is 
\begin{equation}
\mathcal T=\tau_x i s_y K ,
\end{equation} 
where the additional factor $i s_y$ produces the reversal of the electronic spin. It is clear that the mass term $\Delta_{\rm so} \sigma_z \tau_z  s_z$ is now even under $\mathcal T$ because both $\tau_z$ and $s_z$ change signs under time-reversal ($\tau_x \tau_z \tau_x =-\tau_z$ and $s_y s_z s_y = -s_z$). 

Finally we have seen that a perturbation that respects the fundamental symmetries of graphene ($\mathcal T$ and $\mathcal P$) can open a gap owing to the presence of spin. Generally speaking including additional internal degrees of freedom leads to an increase of the number of various possible mass terms. For spinless fermions on the honeycomb lattice, there are only 4 possible mass terms (Semenov, Haldane and two Kekule distortions) and they all break some symmetry. When the spin is included, there are 16 different masses, some of them breaks some symmetries while others, like the Kane-Mele mass, respect all the symmetries \cite{Ryu:2009}.

\medskip

{\it Edge states.} Each subsystem develops a spin-polarized chiral edge state whose circulating direction is tied to the sign of $\phi$ (see chapter \ref{chapter1}). Hence the global Kane-Mele system has a nonchiral edge state consisting in two spin filtered counter propagating gapless edge modes. At the level of the spin-conserving model, those edge states inherit the topological character of the Haldane model edge states. Of course, details like their dispersion relation depend on detail of the surface cut. For instance edge edge states along the armchair cut cross at $k=0$, whereas edge states along the zigzag edge cross at $k=\pm \pi/a$ (projection of the bulk Dirac points). It is a natural question to ask whether this topological character will survive when additional spin-orbit terms mixing the two copies are included. We will see below that the counter propagating edge states are in fact robust as long as the bulk is gapped and time-reversal is obeyed.

From the toy model of the previous chapter (see section \ref{HaldaneSemenov}), one can deduced that each half-part of the QSH state will develop an edge state that circulates clockwise for spin up and anti clockwise for spin down. 

\medskip

{\it Spin Hall conductivity.} Here we can reproduce the Laughlin argument for the QSH state described by Eq.(\ref{KaneMeleHamiltonian}). When a quantum of flux $\phi_0$ is added through the cylinder, one electron is transferred from the bottom to the top of the cylinder for the spin-up subsystem, leading to the $e^2/h$ charge Hall conductivity for this subsystem. For the other subsystem (down-spin), the one electron transfer is realized from the bottom to the top surface of the cylinder. As a result, the Hall charge conductivities cancel out (as required by $\mathcal T$ invariance), but a net spin is transferred from the bottom to the top of the cylinder, namely from one edge to the other. The corresponding spin conductance being:
\begin{equation}
G_{s_z}=\frac{\hbar}{2e} \, . \, \left( \frac{e^2}{h} + \frac{e^2}{h} \right) = \frac{e}{2 \pi} ,
\end{equation}
which is quantized.

Hence, the intrinsic spin orbit coupling Eq.(\ref{HamiltonianKMEffective}) leads to the realization of a new state of matter characterized by an insulating gap and a metallic edge. Nevertheless it is quite unlikely that the spin-orbit coupling manifests itself only through the spin-conserving term above. If this term is present, other terms should be present and mixes the spin component. Since all the previous analysis relies on the description of the system as two decoupled Haldane insulators, the natural question is whether the bulk gap, the helical edge states and the spin Hall effect will survive to the presence of Rashba coupling for instance.

\subsection{Rashba coupling}
Such a Rashba term will be present if the mirror symmetry with respect to the graphene layer is broken (for instance by the presence of a substrate or by a perpendicular electric field). This coupling can be described by the following low energy Hamiltonian \cite{Kane:2005a}:
\begin{equation}
\label{RashbaHamiltonian}
\mathcal{H} _{R}=  \lambda_R \left( \s_x \tau_z s_y - \s_y s_x \right),
\end{equation}
which mixes the two spin directions, and spoils the conservation of $s_z$ because $[\mathcal{H} _{R},\sigma_z] \neq 0$. Typically spin-orbit terms are always described by some coupling of the momentum with the spin Pauli matrices. Here the momentum (for both intrinsic and Rashba spin-orbit couplings) is implicit in the valley degree of freedom $\tau_z$ which encodes the momentum information $\pm \vK$. Note that this term is the lowest order (zero order) possible in the electron momentum $\vq$ measured from the Dirac points. 

\medskip

The corresponding spectrum is \cite{Kane:2005a,Yamakage:2011} 
\begin{equation}
\label{spectrum}
E_{\alpha \beta}(\vq)=  \alpha   \sqrt{\vq^2 + (\Delta_{\rm so} + \beta \lambda_R)^2}  +\beta \lambda_R ,
\end{equation}
where $\alpha=\pm 1$ and $\beta=\pm 1$. As we have seen already the spectrum is gapped in presence of intrinsic spin-orbit coupling only. When the Rashba coupling is increased the gap decreases and the system become gapless when the Rashba coupling exceeds the intrinsic coupling. At $\lambda_R = \Delta_{\rm so}$ the spectrum consists in a Dirac cone and two gapped parabolic bands.

The only limitation is quantitative and is related to the very weak spin-orbit coupling in graphene. The intrinsic SO coupling is extremely small at best 1 mK while Rashba can be around 1K. It has been shown that SO is weak not only because carbon is a light element but also due to the particular arrangement of the $p_z$ orbitals \cite{HuertasPRB:2006,MinPRB:2006}. As a consequence, in order to enhance this intrinsic spin-orbit coupling one should either bend the graphene layer or coat it by heavier atoms like In or .

\section{HgTe/CdTe heterostructures \label{sectionBHZ}}

The Kane-Mele model is a paradigm for 2D topological insulators although the smallness of the intrinsic spin-orbit coupling in graphene hinders the experimental verification. Fortunately we will see that the QSH insulator can be realized in materials with heavier elements and therefore larger spin-orbit coupling. Indeed Bernevig, Hughes and Zhang (BHZ) realized that quantum wells comprising one HgTe layer confined between two barriers of CdTe are good candidates for the realization of the QSH state.  Those authors identified the band inversion between HgTe and CdTe as the crucial ingredient to realize the QSH state. Furthermore a minimal 4 band Dirac model (BHZ model hereafter) captures the essential physics of this band inversion \cite{Bernevig:2006}. This effective model can be derived from the 6 (or 8) band $k.p$ model of 3D HgTe/CdTe complemented by an envelope method to describe the heterostructure. In the absence of spin mixing terms, the QSH insulator realized in this model also consists in two time-reversed copies of a Chern insulator. Since each copy has both a definite spin projection and a definite chirality, the QSH edge states are two spin-filtered counter-propagating modes.   

\subsection{Band structures of HgTe and CdTe}
{\it Three dimensional HgTe and CdTe.} HgTe and CdTe are three-dimensional semimetal and semiconductor respectively. Both HgTe and CdTe crystallize in the zinc-blende band structure which is similar to diamond but with different atoms occupying the two inequivalent sites. HgTe is a semimetal and CdTe a semiconductor. For both HgTe and CdTe, the important bands are close to the center $\Gamma$
of the first Brillouin zone (FBZ). Due to the strong-spin orbit coupling, the total angular momentum $J$ is a good quantum number at least near $\vk =0$. There is a so-called $\Gamma_6$ band made of s-type orbital with spins up or down. There are p-type orbitals which are splitted   into $J=1/2$ (split-off $\Gamma_7$ band) and $J=3/2$ bands ($\Gamma_8$). The $\Gamma_7$ band can be ignored because it is far in energy and it does not participate in the band inversion between the band structures of HgTe and CdTe. At this stage, we have a 6-band model for 3D bulk semiconductors.

\begin{figure}
\includegraphics[width=12cm]{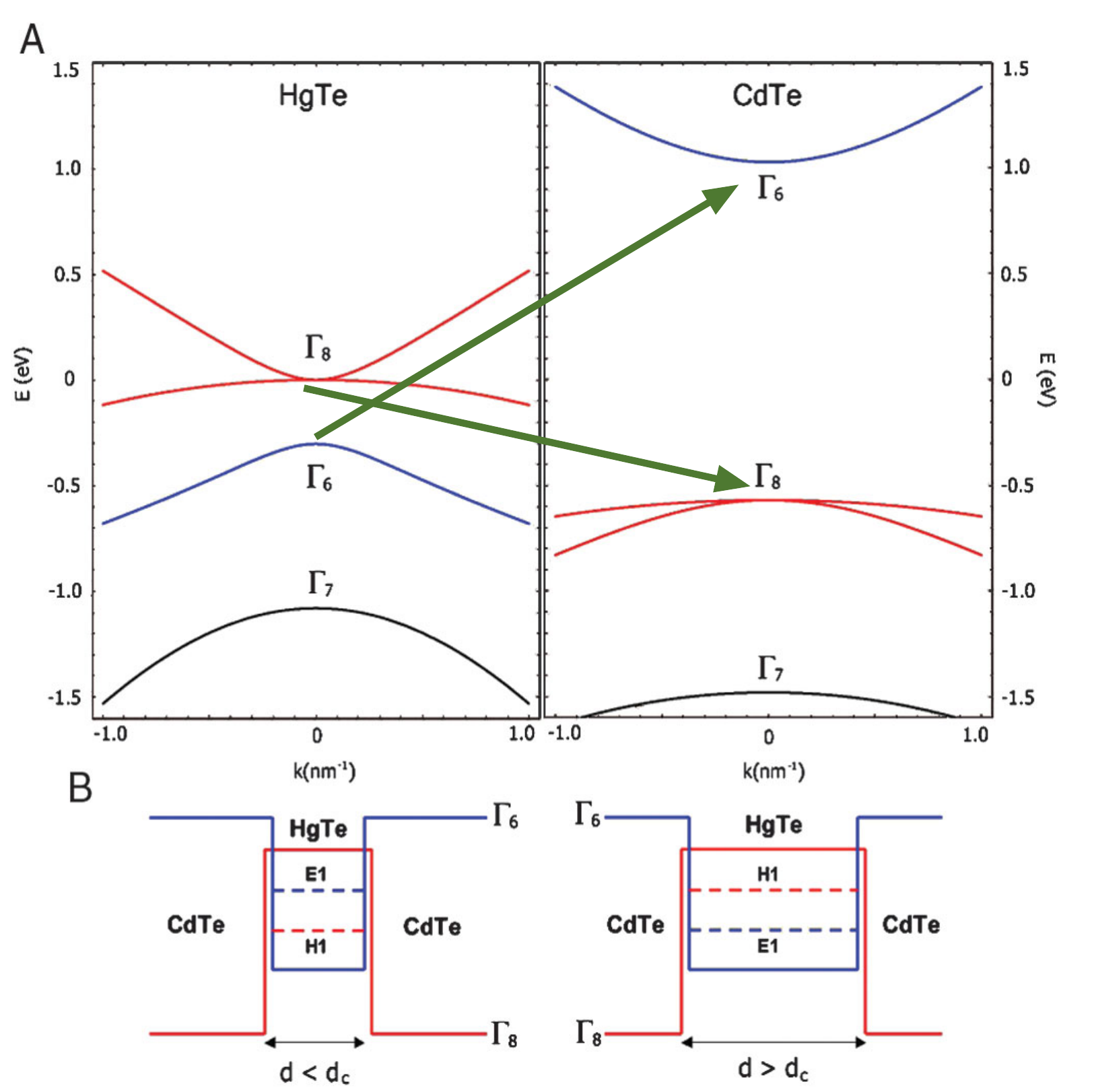}
\caption{Edge state for a single Kramers crossing in the bulk.}
\label{FigTopologyEdge}
\end{figure}

\medskip

{\it Quantum wells.} We now consider a thin HgTe quantum well realized between two identical CdTe
barriers. Using the envelope function method, subband structure of such quantum wells have been
derived from the 6-band Kane model of HgTe and CdTe \cite{Novik2005}. The corresponding 6 states are denoted: $\left\vert E1\pm \right\rangle ,\left\vert
H1 \pm \right\rangle $ and $\left\vert L1 \pm \right\rangle $. It turns out that the degenerate levels electron-like $\left\vert E1\pm \right\rangle$ have a band crossing with the degenerate heavy-hole like $\left\vert
H1 \pm \right\rangle $. In contrast $\left\vert L1 \pm \right\rangle $ does not participate to the crossing, stays far in energy, and therefore can be ignored.

\subsection{Effective model BHZ: massive Dirac fermion}

{\it Continuum version.} The effective model describes the low energy physics near the band-crossing of the 4 states $\left\vert E1\pm \right\rangle$ and $\left\vert H1 \pm \right\rangle $. In the absence of bulk-inversion asymmetry (assuming the atoms "Hg, Te and Cd " were the same atom" thereby transforming the zinc-blende structure in the diamond-like structure), the blocks with opposite spin projections are decoupled. Therefore it is sufficient to study a $2 \time 2$ block, the other block being deduced by TR symmetry. It turns out that the low energy dynamics of this four-band model is captured by the
massive Dirac Hamiltonian \cite{Bernevig:2006} 
\begin{equation}
\mathcal{H}_{4\rm{x}4}(\mathbf{k})=%
\begin{pmatrix}
\mathcal{H}(\mathbf{k}) & 0 \\ 
0 & \mathcal{H}^{\ast }(-\mathbf{k})%
\end{pmatrix}%
,  \label{BHZ4x4}
\end{equation}%
given in the basis order ($\left\vert E1+\right\rangle ,\left\vert
H1+\right\rangle ,\left\vert E1-\right\rangle ,\left\vert H1-\right\rangle $%
). The spin up block:

\begin{equation}
 \mathcal{H}(\mathbf{k})=\epsilon_{0}(\vk)\, {\bold I}_{\rm 2x2}+  \vd(\vk) . \boldsymbol{\sigma} 
 \end{equation} 
is expressed in terms of the standard 
Pauli matrices $\boldsymbol{\sigma}$ acting here in the ($\left\vert E1\right\rangle ,\left\vert
H1\right\rangle $) space. The Hamiltonian $\mathcal{H}(\mathbf{k})$ is a Taylor
expansion with respect to the in-plane wavevector $\mathbf{k=(}k_{x},k_{y})$
whose coefficients are constrained by parity and time-reversal symmetries.

From the symmetries it is possible to infer that the diagonal coupling in the block $\mathcal{H}(\vk)$ are even functions of $\vk$, and the a
diagonal 2$\times$2 matrix can always be decomposed as $\epsilon \boldsymbol{I} + d_z \sigma_z$. In contrast the off-diagonal terms connect opposite parities orbitals and should therefore be odd functions of $k$. Hence one has
\begin{equation}
\vd(\vk) =(A k_x,A k_y,M-B(k_x^2+k_y^2)),
\label{Hamilton2x2BHZ}
\end{equation}
and $\epsilon_0(\vk)=C-D \vk^2$.

Microscopic theory further yields he parameters $A,$ $B,$ $C,$ $D$ and $M$ as functions of
the quantum well geometry \cite{Bernevig:2006}. The parameters $A \simeq - 3.8 $eV.A and $B \simeq - 60 $ eV.A$^2$ have a definite sign over the relevant range of thicknesses, whereas $M$ changes sign at roughly $d=6.4$ nm.

The inversion between $\left\vert E1\right\rangle $ and $\left\vert
H1\right\rangle $ is controlled by the sign of the mass term $M$, the QSH
state being realized in the inverted regime ($M<0$). In fact the trivial/non trivial topology of 
this two-band model (spin up block) is determined by the relative signs of $B$ and $M$. Qualitatively, this can be inferred from 
the component $d_z(\vk)=M-B\vk^2$ which indicates that $MB$ is positive the vector $\vd$ points to opposite directions (north and south poles) at the $\Gamma$ point (\vk=0) and far from it (large $|\vk|$). This configuration which looks like a meron in $\vk-$space is nontrivial. In contrast if $MB<0$, then the texture $\vd(\vk)$ has no winding. Since material parameters lead to $B<0$, the nontrivial regime corresponds to the inverted regime $M<0$.     

The chemical potential $%
C $ determines the electronic filling of the bands which can be
electrostatically tuned by the action of a distant metallic gate.

\medskip

{\it Regularized lattice version.} 
The above model is already simplified and has the advantage of being quantitative (values of parameters derived from a more microscopic theory). Nevertheless being a $k.p$ expansion it is only valid locally near $\vk=0$. Of course one cannot solve the full band structure in the whole FBZ, but one can imagine a simple idealized model which is defined over the whole FBZ and has the same low $k$ expansion as effective model Eq. (\ref{Hamilton2x2BHZ}), namely:
\begin{equation}
\vd(\vk) =(A \sin k_x,A \sin k_y,M-2B(2-\cos k_x-\cos k_y)).
\label{Hamilton2x2TBAk}
\end{equation}
This can be represented as a square lattice with four states at each site: $\psi_{1,3}=s$ is s-type orbital with up/down spin, while $\psi_{2,4}=p_x \pm i p_y$-spin orbit coupled orbital with up/down spin. \medskip

\subsection{Edge states of the BHZ model: interface with the vacuum }
We now investigate the edge states of HgTe/CdTe quantum wells using the BHZ model which neglect the spin-mixing BIA and SIA terms. Then it is sufficient to consider the spin-up block $\mathcal{H}(\vk)$ in Eq.(\ref{BHZ4x4}). For simplification, we neglect the $\epsilon_0 Id$ term which is not relevant for the topological properties. 

We consider an interface along $y-$axis between the HgTe/CdTe well ($x>0$) and vacuum ($x<0$). Owing to translational invariance along $y-$axis, the momentum $k_y$ is a good quantum number whereas $k_x$ should be replaced by the real-space derivative $\partial_x$. In the half-plane $x>0$, the wave function then obeys a sec ond-order differential equation:
\begin{equation}
\left( -i A \sigma_x \partial_x + A k_y  \sigma_y + (M + B \partial_x^2 -B k_y^2) \sigma_z  \right) \Psi = E \Psi,
\label{waveeqBHZvacuum}
\end{equation}
with the strict boundary condition $\Psi(x=0)=0$. We search for a zero energy mode at $k_y=0$. Hence we start by solving Eq.(\ref{waveeqBHZvacuum}) with $k_y=0$ and $E=0$: 
\begin{equation}
\left( -i A \sigma_x \partial_x  + (M + B \partial_x^2 ) \sigma_z  \right) \Psi = 0,
\label{waveeqBHZvacuumkyzero}
\end{equation}
which can also be written as:
\begin{equation}
 - B \partial_x^2  \Psi - A  \sigma_y \partial_x \Psi   =  M \Psi.
\label{waveeqBHZvacuumkyzerobis}
\end{equation}
If we inject an exponential solution $\Psi=\phi e^{\lambda x}$, one finds 
\begin{equation}
(M + B \lambda^2)  \Phi_{\sigma} = - A  \sigma_y \lambda \Phi_\sigma,
\label{waveeqPhi}
\end{equation}
and therefore the solution are eigenmodes of $\sigma_y$ with decay lengths $\lambda^{-1}$ given by:
\begin{equation}
\lambda_{\pm}^{\sigma_y} = \frac{- A \sigma_y \pm \sqrt{A^2 -4 BM}}{2B},
\label{lambda}
\end{equation}
where in this equation $\sigma_y$ has to be understood as an index $\sigma_y=\pm$ referring to the eigenvalue of the matrix $\sigma_y$. Then if we solve the problem in the half-plane $x>0$, one should pick up the solutions with Re$\lambda <0$ which have physical asymptotic behavior at $x \rightarrow \infty$. Then two scenarios are possible. If two solutions have the same $\sigma_y$-polarization, then it is possible to satisfy the condition $\Psi(x=0)=0$ and the solution reads:  
\begin{equation}
\Psi(x) = \Phi_{\sigma_y} (e^{\lambda_1 x} - e^{\lambda_2 x}),
\label{solution}
\end{equation}
where $\sigma_y$ is the chosen polarization. Since $\lambda_{+}^{\sigma_y} .\lambda_{-}^{\sigma_y}=M/B$, this situation is realized if $M/B>0$. In contrast, when $M/B<0$, the two solutions having the right asymptotic behavior belongs to orthogonal polarizations, and it is impossible to cancel the wave function at $x=0$ by the superposition of two orthogonal spinors.

In terms of the parameters, the condition for having an edge state turns out to be $A^2 > 4 MB >0$. The signs $A >0$ and $B<0$ are fixed over a broad range of HgTe width $d$, whereas the "mass" $M$ changes it sign as a function at $d_c = 64$ nm. According to the BHZ model, there is a zero mode $E=0$ at $k_y=0$ when $M<0$.

{\it Dispersion and Hamiltonian of the edge state.} For $M<0$, we can restore a finite $k_y$ and notice that the zero energy solution is still an eigensolution at energy $E=\sigma_y A k_y $ if one neglects the terms $B k^2 \sigma_z$ (which is always possible at very low $k_y$). The $\sigma_y =\pm$ indicates the polarization of such an edge mode. Note that such an edge mode is chiral which was expected because it is the edge mode of the Chern insulator described by the spin-up block of the BHZ model. The full model has an edge mode carrying opposite spin, opposite $\sigma_y$ polarization and circulating in the opposite direction.

\subsection{Interface between two BHZ insulators with different masses}

Another model we may want to solve is the interface between two BHZ insulators characterized by distinct values of the mass. The corresponding equation being:
\begin{equation}
\left( -i A \sigma_x \partial_x + A k_y  \sigma_y + (M(x) + B \partial_x^2 -B k_y^2) \sigma_z  \right) \Psi = E \Psi,
\label{waveeqTwoBHZ}
\end{equation}
with $M(x)=M_1 \Theta(-x) + M_2 \Theta(x)$.

Now we can try to construct the edge states solutions for a semi-infinite
HgTe well extending in the half-plane $y>0$. Boundness requires to choose
the two exponential with negative $\lambda $ while the boundary condition is 
$\Psi (x,y=0)=0$. Note that in the particular model of HgTe wells near the
transition point, parameters $A$ and $B$ are both negative. Thus $\lambda
_{1}$ is always negative whereas the sign of $\lambda _{2}$ is the sign of $%
-MB$. We now discuss the cases $MB>0$\textbf{\ }and $MB<0$.

\textbf{Conclusion: We find no edge state between two nontrivial (or two
trivial) insulators. We find a gapless edge state between a trivial and
a nontrivial insulators. Note that the edge state consists in a single
exponential on the trivial side while it is characterized by two
exponentials on the nontrivial side.}

\section{Helical edge states \label{sectionedge}}

In the previous examples of spin conserving models, we have seen that QSH insulators consist in two copies of a Chern insulator (or QAH state), each copy being associated with a spin orientation. Since the QAH has (minimaly) a single chiral edge state, we can deduce that the QSH state will have two spin-filtered counter propagating states. Due to spin-conservation, there is no backscattering coupling those states. Nevertheless one may question if the edge states pertain for more realistic models that include the spin-mixing unavoidably present in any material with stron spin-orbit coupling.

The crucial question is the robustness of this helical state in presence of spin mixing. Numerical indications that the edge states are preserved as long as the bulk is gapped. This paragraph tries to explain the general reason for this protection which is Kramers degeneracy.
We aim at describing the implications of Kramers degeneracy for Bloch electronic band structures, both at the level of eventual edge states and at the bulk level. Finally the name quantum spin insulator can be misleading. This is a new insulating state that is different from the doped semiconductors exhibiting QSH effect (metals versus insulators). Nevertheless if the Fermi level is raised into the 2D conduction or valence bands, we obtain such states.

\subsection{Kramers degeneracy}
The time-reversal operation $\mathcal T$ is anti-unitary. Then $\mathcal T^2 =\pm 1$ depending on the spin of the system. We are interested in fermionic systems, and then $\mathcal T^2 =-1$. Then one can prove that for a time-reversal invariant system, characterized by $[H,T]=0$, any state is always degenerated. Indeed if $\phi$ is a stationary state with energy $E$, namely $H\phi=E\phi$, then $\mathcal{T}\phi$ is also an eigenstate of $H$ with the same eigenvalue. Moreover $\mathcal{T}\phi$ is a different quantum state than $\phi$ precisely because of $\mathcal T^2 =-1$. Indeed if $\mathcal{T}\phi$ and $\phi$ were the same state, then one could find a complex number $c$ realizing $\mathcal{T}\phi = c \phi$ and therefore $\mathcal{T}^2 \phi = c^{*} \mathcal{T}\phi=|c|^2 \phi$. The last equality leads to the contradiction $|c|^2=-1$ because $\mathcal{T}^2=-1$.

\subsection{Configuration of edge states}
The spirit of this section is the following. Let us assume the existence of some 1D edge states within the bulk gap. Then we shall examine the implications of time-reversal symmetry on the dispersion of such edge states $\epsilon(k)$. The dispersion is defined with respect to a momentum parallel to the border of the insulator, denoted $k$, which lives in the segment $[-\pi/a,\pi/a]$. The time-reversal invariant points are $k=0$ and $k=\pi/a$ (which is the same as $k=-\pi/a$).

\begin{figure}
\includegraphics[width=6cm]{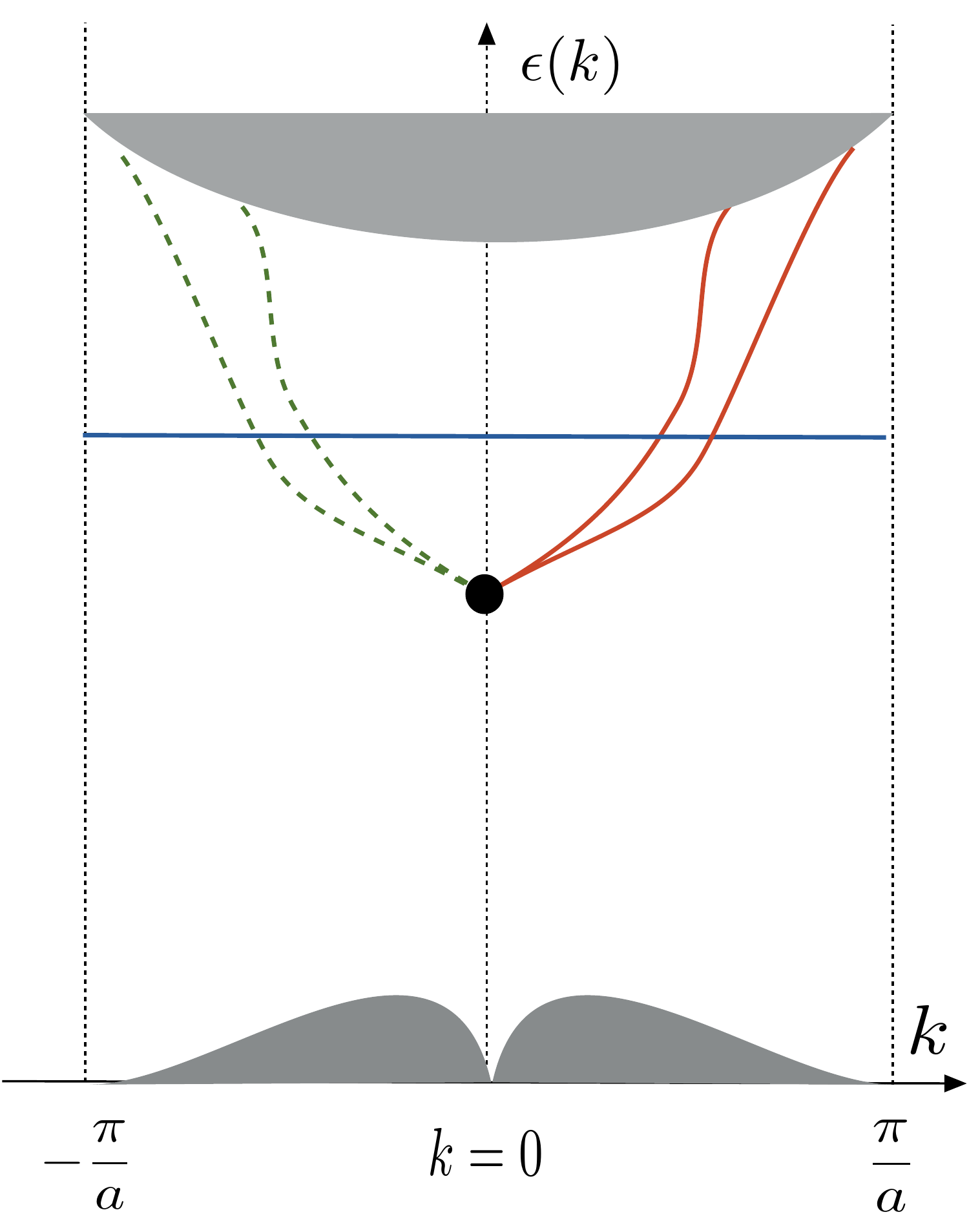}
\includegraphics[width=6cm]{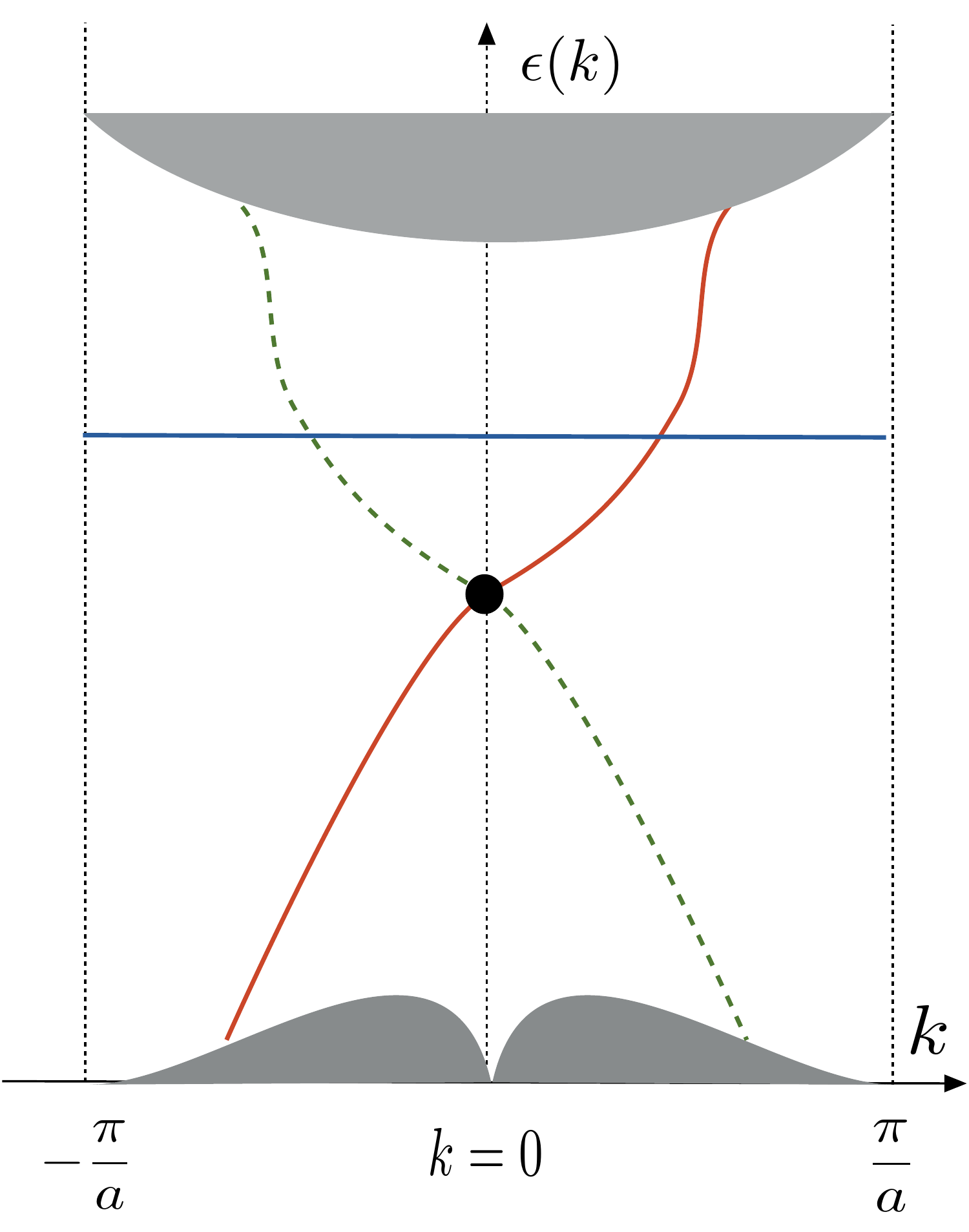}
\caption{Edge state for a single Kramers crossing in the bulk.}
\label{FigTopologyEdge}
\end{figure}

{\it A single Kramers doublet.} The minimal situation corresponds to 1 single Kramers doublet forming at the TRI points that we can take at $k=0$ for simplicity as in Fig. \ref{FigTopologyEdge}. When $k$ is varied from $0$ to $\pi/a$ say, the degeneracy is lifted and the two levels can move inside the bulk gap. Assuming a single Kramers doublet for this surface state, the splitted Kramers partners have to merge into the bulk bands either the conduction or valence bands.Then we have two possible scenarios depending whether the Kramers partners merge two the same bulk band or to distinct bands.

\begin{figure}
\begin{center}
\includegraphics[width=6cm]{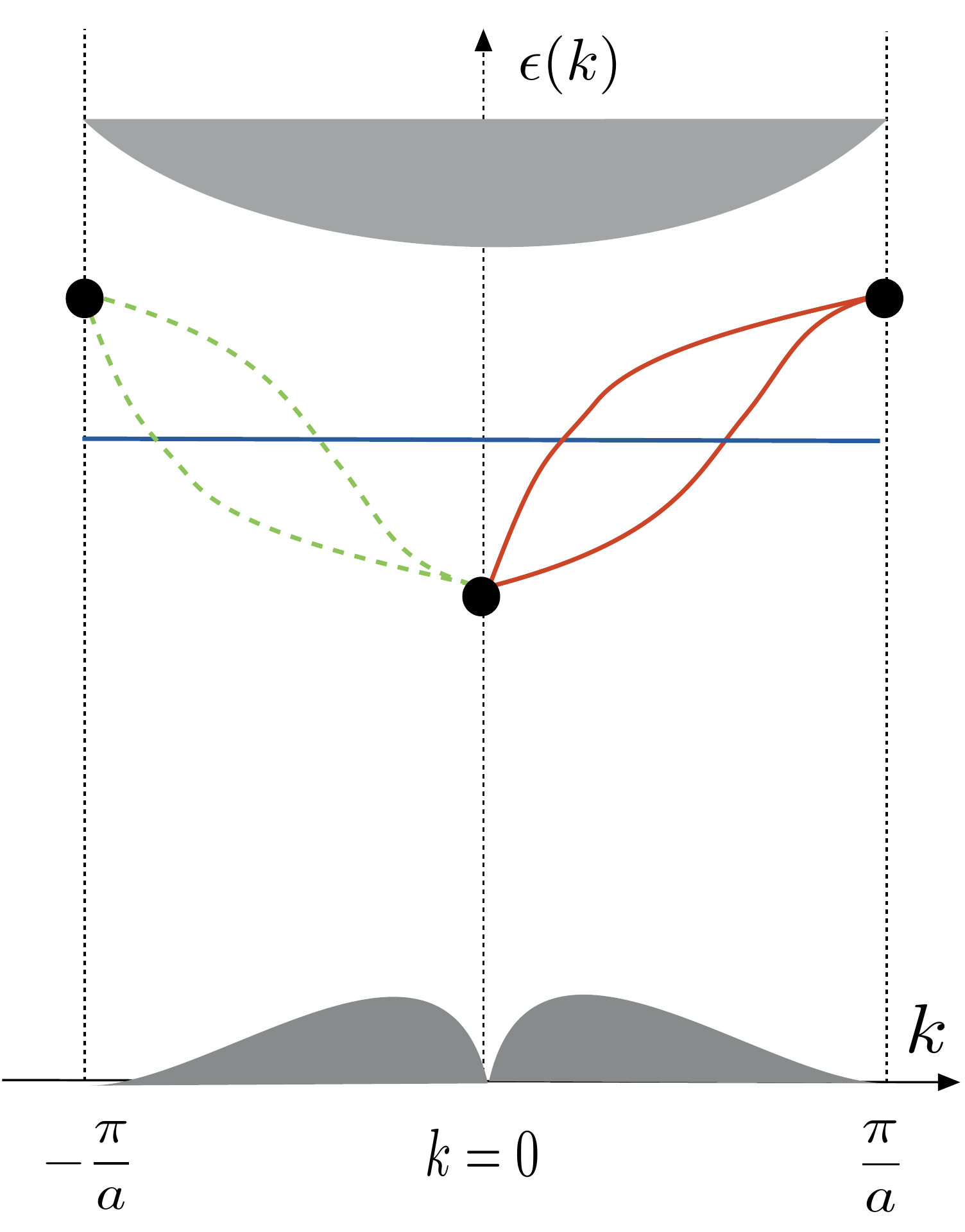}
\includegraphics[width=6cm]{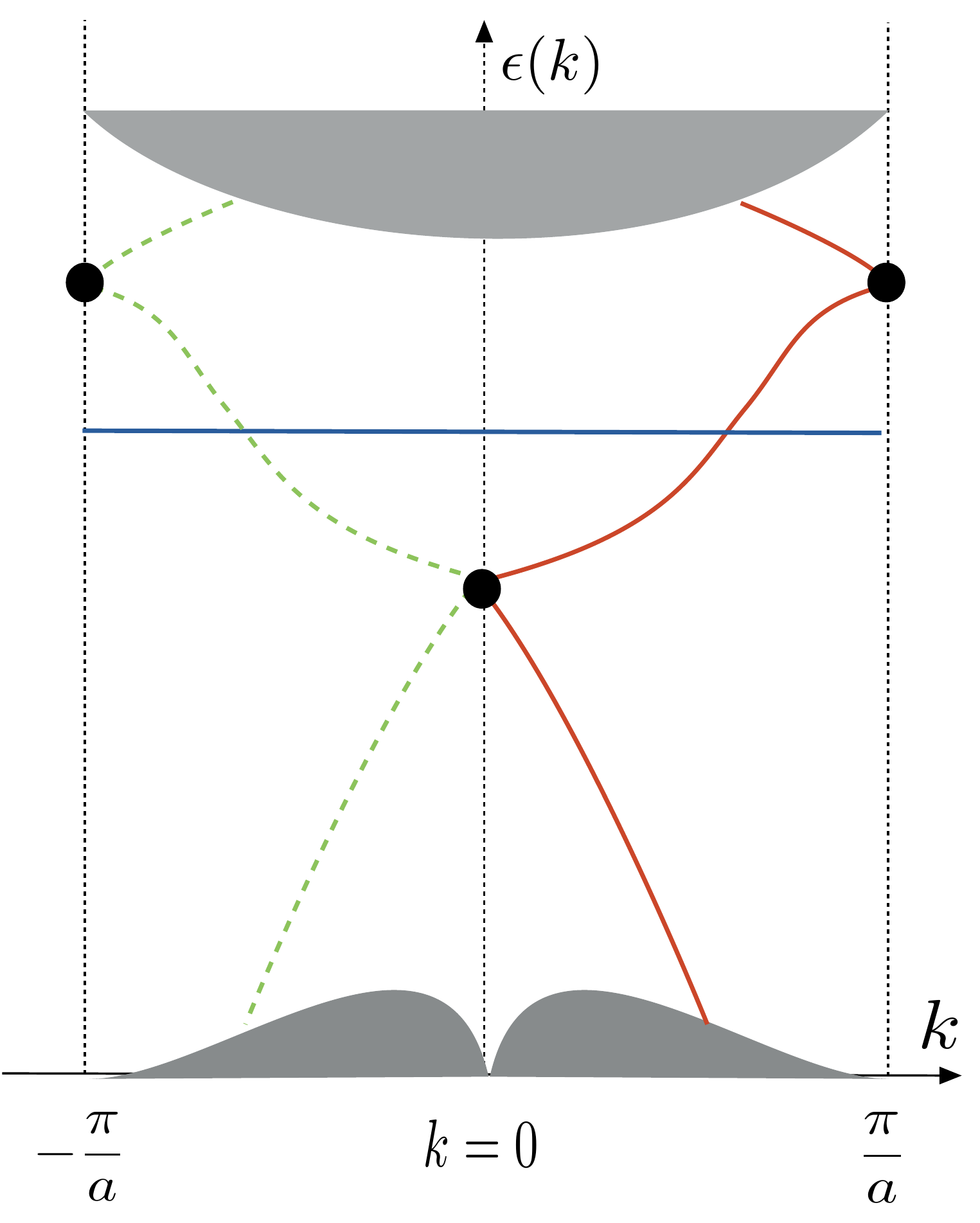}
\caption{Edge state for two Kramers crossing in the bulk.}
\end{center}
\label{FigTopology2Edges}
\end{figure}

{\it Two Kramers doublets.} We now imagine a slightly more complex situation with Kramers doublets at $k=0$ and $k=\pi/a$ within the bulk gap. We also assume that the states belonging to those two doublets will become intricate in some way (otherwise we are left with the previous situation of two independent decoupled Kramers pairs).
Then there are two possible scenarios illustrated in Fig. \ref{FigTopology2Edges}. First, the two partners can stay in the gap region and merge again together at $k=\pi/a$ (Fig. \ref{FigTopology2Edges}.a). Second, one of the partner can merge into the bulk valence band while the other will stay away from the band edge and will be joined by a state arising from the conduction band (\ref{FigTopology2Edges}.b). 
In the first case, the Fermi level in the gap will cross either two or zero pair(s) of edge states. This is similar to a standard 1D conductor. Moreover if the Kramers degeneracy at TRI points are moved both up to conduction band (or both down to valence band) the edge states disappear. 

In the second case, the FL intersects only a single pair. The two counter propagating modes at the Fermi level are each non degenerate and carry a given Kramers parity. This is half of a usual 1D system. Moreover if we move the two Kramers doublet to the bulk bands (in any way), there will always remain a pair of Kramers partners crossing the bulk gap.
 
\subsection{Absence of single-particle backscattering}
We have seen that the crossing of the time-reversed states is protected by $\mathcal T$ at the time-reversal invariant points of momentum space. Now we investigate the relation between time-reversed states, typically labelled by $\vk$ and $-\vk$, at any momentum $\vk$. It can be proved that single-particle processes between such states are forbidden. This extends dramatically the absence of backscattering discussed in the previous sections on spin-conserving models. 

This can be also proved in a very general way general way. Let us consider two time-reversed states $\phi$ and $\psi =\mathcal{T} \phi$. Then we can show that the matrix element $\langle \psi | H \phi  \rangle$ is always zero provided the system is time-reversal invariant $[H,\mathcal{T}]=0$ and fermionic $\mathcal{T}^2=-1$:

\begin{align}
\langle \psi | H \phi  \rangle &= \langle \mathcal{T} \phi | H \phi  \rangle=\langle \mathcal{T}H \phi | \mathcal{T}^2 \phi  \rangle \\
& = - \langle \mathcal{T}H \phi |  \phi  \rangle  \,   \rm{using} \,  \mathcal{T}^2=-1\\
& = - \langle H \mathcal{T} \phi |  \phi  \rangle  \,   \rm{using} \,   from [\mathcal{T},H]=0  \\
& = - \langle \mathcal{T} \phi | H |  \phi  \rangle
\end{align}
There is another proof using second-quantization and based on the transformation rule of creation/annihilation operators for one-half particles \cite{QiRMP:2011}:
\begin{align}
\mathcal{T} c_{\vk \uparrow} \mathcal{T}^{-1}  &= c_{-\vk \downarrow} \\
\mathcal{T} c_{-\vk \downarrow} \mathcal{T}^{-1}  &=-  c_{\vk \uparrow} 
\end{align}
Due to those properties the Hamiltonian describing single-particle backscattering for a single pair of Kramers partners \cite{QiRMP:2011}:
\begin{equation}
H_{\rm back}= \sum_{\vk,s}  c_{\vk s}^\dagger c_{-\vk -s} + H.c.,
\end{equation}
is odd under $\mathcal T$:
\begin{equation}
\mathcal{T} H_{\rm back} \mathcal{T}^{-1}=  - H_{\rm back}.
\end{equation}

Finally, we note that two-particle processes are allowed. Recently it was discussed that they can be generated microscopically by
the interplay of Rashba coupling and electron-electron interactions.

\subsection{Z$_2$ topological invariant \label{sectionbulk}}

The central idea is to implement the condition of time-reversal invariance in the context of systems where spin-orbit induces a band inversion and a related topological state. If the edge states are present, then TRI enforces that their dispersion curves should meet at the TRI points of the projected BZ. The bulk states should also be constrained by time-reversal symmetry. When TR is obeyed the TKKN invariant is zero. Kane and Mele shown that there is an other invariant which can be $\nu=0,1$ only. There is a formula for this $Z_2$ topological invariant which can be expressed as an integral over the whole bulk BZ. For systems having a center of inversion, the additional $\mathcal P$ symmetry allows to simplify this formula and to calculate the topological invariant $\nu=0$ has a simple product of the parities of occupied states at all the inequivalent TRI points of the FBZ.  

In the QH effect (and in Chern insulators) there is a correspondence between the value of the bulk topological invariant of the occupied bulk bands (the TKKN or Chern number) and the number of chiral edge states. By qualitative considerations we have found that there should be a $Z_2$ classification of the edge states of time-reversal invariant insulators related to the even/odd effect in the number of edge modes crossing the Fermi level. In presence of inversion symmetry, there is a simplified formula to evaluate the Z$_2$ invariant  $\nu$ and determine whether a given band structure is a nontrivial topological insulator or a trivial band insulator. The formula is simply \cite{LiangFu:2007}:
\begin{equation}
(-1)^\nu =\prod_{\vk_{\rm inv}}   \prod_{m=1,..,N}  \xi_{2m},
\end{equation}
where $\xi_{2m}=\xi_{2m+1}$ is the parity of the Bloch wave function at $\vk=\vk_{\rm inv}$ in the occupied band labelled by $2m$.

\section*{Conclusion}

The Quantum Spin Hall (QSH) effect is a property of certain two-dimensional electron systems with spin-orbit coupling: the bulk of the system is electrically insulating, while a conducting "helical metal" exists at the boundary in which electrons of opposite spins move in opposite directions \cite{Kane:2005a,Kane:2005b}. The prediction \cite{Bernevig:2006} and the observation \cite{Konig:2007,Roth:2009} of the Quantum Spin Hall (QSH) state in mercury telluride (HgTe/CdTe) heterostructures have triggered a great deal of excitation in the condensed matter community \cite{KaneRMP:2011,QiRMP:2011,QiPhysToday:2010} since the QSH state realizes a two dimensional (2D) topologically ordered phase in the absence of magnetic field.

Finally we mention that the QAH and QSH phases might be generated spontaneously by some types of local Hubbard correlations \cite{Raghu:2008,Rachel:2010} even in material with light elements with weak spin-orbit couplings in the absence of interactions.








\appendix 


\bibliographystyle{unsrt}

\cleardoublepage
\ifdefined\phantomsection
  \phantomsection  
\else
\fi
\addcontentsline{toc}{chapter}{Bibliography}

\bibliography{HdrThesis}



\end{document}